\documentclass[apj, 8pt]{emulateapj-rtx4}
\usepackage{color}
\usepackage{amsmath}

\defcitealias{2015ApJS..219...15S}{Paper I}
\defcitealias{2016ApJ...821...72S}{Paper II}

\usepackage[dvipdfm]{hyperref}
\hypersetup{
	bookmarks=true,
	bookmarksnumbered=true,
	bookmarksopen=true,
	colorlinks   = true,
	citecolor    = blue,
	pdftitle = {Continuum Profile and Size Evolution of LAEs},
	pdfauthor = {Takatoshi Shibuya},
	pdfkeywords={cosmology: observations --- early universe --- galaxies: formation --- galaxies: high-redshift}
	pdfpagemode=UseThumbs
}

\renewcommand\thefootnote{\arabic{footnote}}

\slugcomment{Submitted to ApJ, 2018 September 4; accepted 2018 December 3}

\shorttitle{Continuum Profile and Size Evolution of LAEs}
\shortauthors{T. Shibuya et al.}

\begin{document}

\title{Morphologies of $\sim 190,000$ Galaxies at $\lowercase{z}=0-10$ Revealed with HST Legacy Data. III. \\
Continuum Profile and Size Evolution of Ly$\alpha$ Emitters}

\author{Takatoshi Shibuya\altaffilmark{1, 2}, Masami Ouchi\altaffilmark{2, 3}, Yuichi Harikane\altaffilmark{2, 4}, and Kimihiko Nakajima\altaffilmark{5}}
\email{tshibuya\_@\_mail.kitami-it.ac.jp}

\altaffiltext{1}{Kitami Institute of Technology, 165 Koen-cho, Kitami, Hokkaido 090-8507, Japan}
\altaffiltext{2}{Institute for Cosmic Ray Research, The University of Tokyo, 5-1-5 Kashiwanoha, Kashiwa, Chiba 277-8582, Japan}
\altaffiltext{3}{Kavli Institute for the Physics and Mathematics of the Universe (Kavli IPMU, WPI), University of Tokyo, Kashiwa, Chiba 277-8583, Japan}
\altaffiltext{4}{Department of Physics, Graduate School of Science, The University of Tokyo, 7-3-1 Hongo, Bunkyo, Tokyo, 113-0033, Japan}
\altaffiltext{5}{National Astronomical Observatory of Japan, 2-21-1 Osawa, Mitaka, Tokyo 181-8588, Japan}

\begin{abstract}

We present the redshift evolution of the radial surface brightness (SB) profile of the rest-frame UV and optical stellar continua for $9119$ Ly$\alpha$ emitters (LAEs) at $z\simeq0-8$ and $0-2$, respectively. Using $\!${\it Hubble Space Telescope} data and the LAE catalogs taken from the literature, we derive the structural quantities of the $9119$ LAEs and $\simeq180,000$ comparison galaxies of photo-$z$ star-forming galaxies (SFGs) and Lyman break galaxies (LBGs) by the well-tested profile fitting. From 936 well-fitted LAEs, we carefully define the homogeneous sample of LAEs falling in the same ranges of UV-continuum luminosity and Ly$\alpha$ equivalent width over $z\simeq0-8$, and evaluate the redshift evolution. We find that the effective radius $r_{\rm e}$ distribution is represented by a log-normal function, and that the median S\'ersic index is almost constant at $n\simeq1-1.5$ for the LAEs over $z\simeq0-7$, suggesting that typical LAEs have a stellar-disk morphology. The size-luminosity relation of the LAEs monotonically decreases towards high-$z$, following size-luminosity relations of SFGs and LBGs. The median $r_{\rm e}$ values of the LAEs significantly evolve as $r_{\rm e}\propto(1+z)^{-1.37}$, similar to those of the SFGs and LBGs in the same luminosity range, in contrast with the claims of no evolution made by previous studies whose LAE samples are probably biased to faint sources at low-$z$. The $r_{\rm e}$ distribution, star-formation rate surface densities, and stellar-to-halo size ratios of the LAEs are comparable with those of the SFGs and LBGs, indicating that LAEs have stellar components similar to SFGs and LBGs with a Ly$\alpha$ emissivity controlled by the non-stellar physics such as geometry, kinematics, and ionization states of the inter-stellar/circum-galactic medium.

\end{abstract}


\section{INTRODUCTION}

\begin{figure}[t!]
  \begin{center}
    \includegraphics[width=90mm]{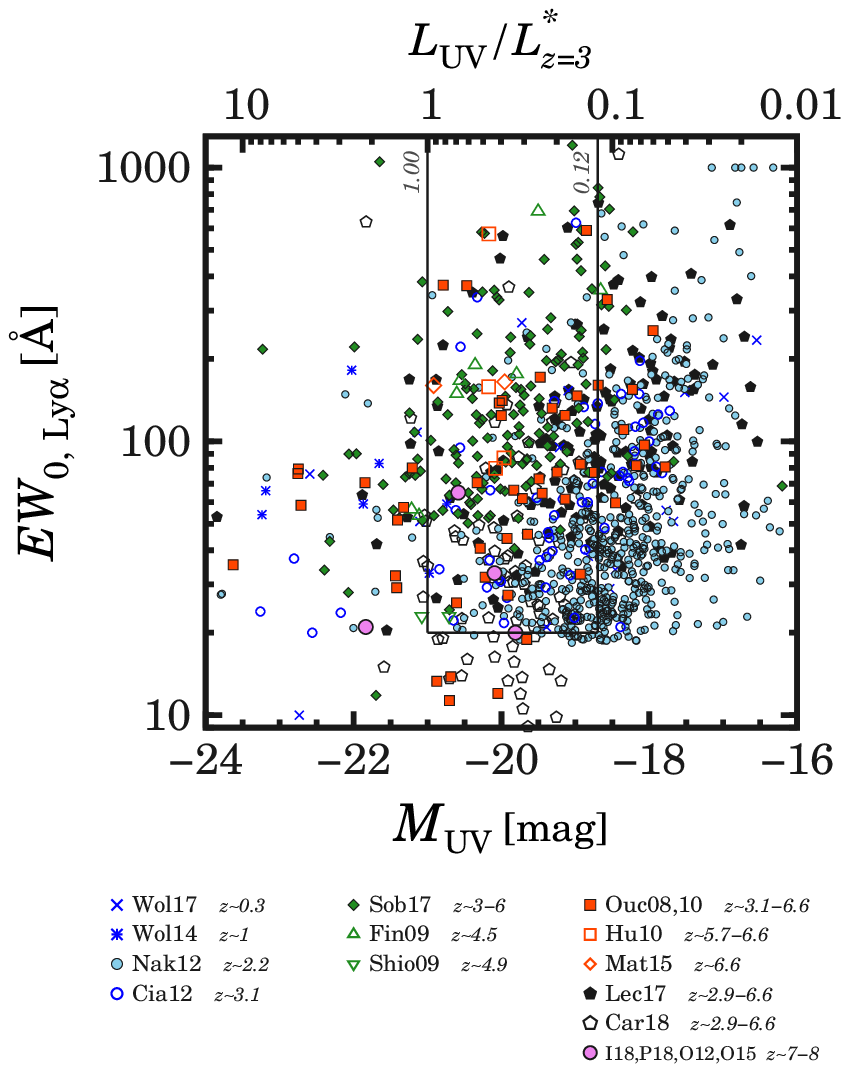} 
  \end{center}
  \caption[]{{\footnotesize Ly$\alpha$ EW as a function of UV magnitude for LAEs used in this study (blue crosses: \citealt{2017ApJ...848..108W}; blue asterisks: \citealt{2014ApJ...783..119W}; cyan filled circles: \citealt{2012ApJ...745...12N}; blue open circles: \citealt{2012ApJ...744..110C}; green filled diamonds: \citealt{2017arXiv171204451S}; green open triangles: \citealt{2009ApJ...691..465F}; green open inverse-triangles: \citealt{2009ApJ...700..899S}; red filled squares: \citealt{2008ApJS..176..301O, 2010ApJ...723..869O}; red open squares: \citealt{2010ApJ...725..394H}; red open diamonds: \citealt{2015MNRAS.451..400M}; black filled pentagons: \citealt{2017AA...608A...8L}; black open pentagons: \citealt{2018MNRAS.473...30C}; magenta filled circles; \citealt{2018arXiv180505944I}, \citealt{2018arXiv180801847P}, \citealt{2012ApJ...744...83O}, and \citealt{2015ApJ...804L..30O}). The Ly$\alpha$ EW of \citet{2018arXiv180505944I}'s LAE is assumed to be $EW_{\rm 0,Ly\alpha}=20$\,\AA. This diagram shows LAEs whose effective radius is obtained in our {\tt GALFIT} measurements. The top x-axis provides the corresponding UV luminosity in units of $L_{z=3}^*$ (see Section \ref{sec_sample}). The vertical and horizontal lines denote the thresholds of UV luminosity (i.e., $L_{\rm UV}=0.12 - 1 L_{z=3}^*$) and Ly$\alpha$ EW (i.e., $EW_{\rm 0, Ly\alpha} > 20$\,\AA), respectively, for our analysis. }}
  \label{fig_muv_ewlya}
\end{figure}

\begin{figure}[t!]
  \begin{center}
    \includegraphics[width=90mm]{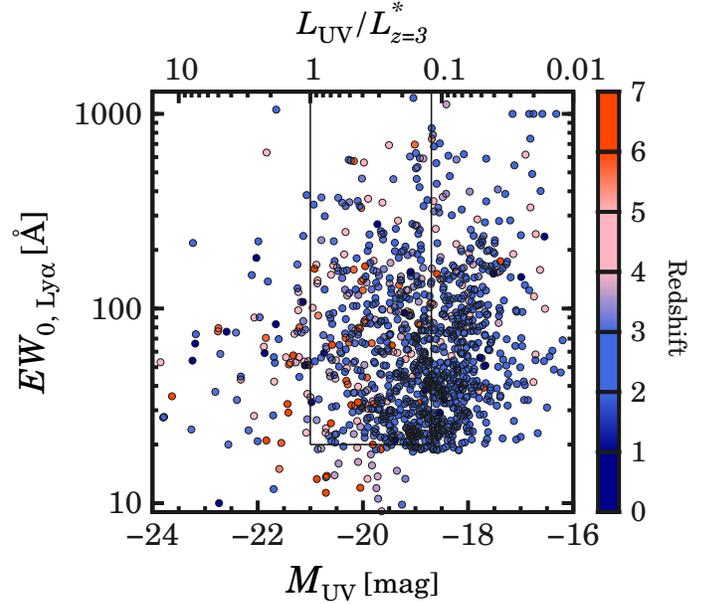}
  \end{center}
  \caption[]{{\footnotesize Same as Figure \ref{fig_muv_ewlya}, but for LAEs that are color-coded based on the redshift.}}
  \label{fig_muv_ewlya_z}
\end{figure}

Galaxy morphological properties provide us with invaluable hints for understanding the galaxy formation and evolution mechanisms. Particularly, radial surface brightness (SB) profiles and angular sizes of galaxies contain a lot of crucial information about e.g., galaxy build-up processes and the relation between stellar components and underlying dark matter (DM) halos. These key morphological quantities are studied extensively with high spatial resolution images of the $\!${\it Hubble Space Telescope} ($\!${\it HST}) for star-forming galaxies (SFGs), quiescent galaxies \citep[e.g., ][]{2003MNRAS.343..978S, 2014ApJ...788...28V} and Lyman break galaxies (LBGs) up to redshift $z\simeq10$ (e.g., \citealt{2004ApJ...611L...1B, 2012AA...547A..51G,2013ApJ...765...68H, 2013MNRAS.428.1088M, 2013ApJ...777..155O, 2015ApJS..219...15S, 2015ApJ...808....6H, 2017ApJ...843...41B}; see also \citealt{2018arXiv180909136D} for a compilation of size evolution studies). Over the past decades, deep survey data of $\!${\it HST} have allowed to uncover a variety of the morphological nature not only for such relatively massive populations, but also for a typically low-mass system of Ly$\alpha$ emitters \citep[LAEs; e.g., ][]{2005AA...431..793V, 2007ApJ...667...49P, 2008ApJ...673..143O, 2009ApJ...705..639B, 2009ApJ...701..915T, 2011ApJ...743....9G,  2012ApJ...750L..36M, 2012ApJ...753...95B, 2013ApJ...773..153J, 2014ApJ...782....6H, 2014ApJ...785...64S, 2015AA...576A..51G, 2016ApJ...817...79H, 2016ApJ...819...25K, 2016AA...587A..98W, 2017AA...608A...8L, 2017MNRAS.468.1123S, 2017ApJ...838....4Y, 2018MNRAS.476.5479P}. 

Some of the previous studies have quantified the radial SB profiles of the stellar continuum emission for LAEs with the S\'ersic index \citep{1963BAAA....6...41S,1968adga.book.....S} to be $n\simeq1$ at $z\simeq3.1$ \citep{2011ApJ...743....9G} and $z\simeq5.7$ \citep{2009ApJ...701..915T}. The results of the low $n$ values suggest that LAEs have a disk-like SB profile. Recently, \citet{2018MNRAS.476.5479P} have shown that LAEs have a nearly constant S\'ersic index of $n\lesssim2$ at $z\simeq2-6$ with a large sample of $\simeq3000$ LAEs, indicating that the disk-like SB profile of LAEs does not significantly change over the cosmic time. In addition to the constraints on $n$, the radial SB profiles are useful for investigating the origins of extended Ly$\alpha$ emission, so-called Ly$\alpha$ halos, surrounding high-$z$ star-forming galaxies \citep[e.g., ][]{2004AJ....128.2073H, 2008ApJ...681..856R, 2011ApJ...736..160S, 2012MNRAS.425..878M, 2014MNRAS.442..110M, 2016MNRAS.457.2318M, 2016MNRAS.458..449M, 2016AA...587A..98W, 2017MNRAS.466.1242S, 2017ApJ...837..172X, 2017AA...608A...8L}. The diffuse and extended emission of the Ly$\alpha$ halos is expected to result from e.g., cold streams accreted onto the central galaxies, resonantly scattered Ly$\alpha$ radiation produced by star formation activities of neighboring small satellite galaxies, and/or the fluorescent Ly$\alpha$ emission created by the ionizing photons from the central sources \citep[see recent review papers, e.g., ][]{2017arXiv170403416D, 2018arXiv180909136D}. The image stacking analysis of ground-based telescope data has detected the faint and diffuse Ly$\alpha$ emission down to SB $\simeq 10^{-32-(-33)}$ erg$^{-1}$ s$^{-1}$ cm$^{-2}$ Hz$^{-1}$. Complementary to the image stacking analysis, the $\!${\it HST} high spatial resolution observations enable us to compare the radial SB profiles in Ly$\alpha$ and the stellar continuum radiation near the galaxy center. The central part of the stellar radial SB profiles would be used for modeling star formation budgets from small satellite galaxies as a function of angular distance. 

On the other hand, measurements of galaxy sizes, defined as the effective radius, $r_{\rm e}$, in the stellar continuum emission have shown that LAEs are compact and do not typically evolve at $r_{\rm e}\simeq1$ kpc in the redshift range of $z\simeq2-6$ \citep{2012ApJ...750L..36M, 2014ApJ...786...59H, 2018MNRAS.476.5479P}. For example, \citet{2018MNRAS.476.5479P} report that the median $r_{\rm e}$ scales as $r_{\rm e} \propto (1+z)^{-0.21\pm0.22}$ at $z\simeq2-6$ for LAEs, which is consistent with no size evolution scenario within a $1\sigma$ uncertainty. This is in contrast to the size evolution with $r_{\rm e} \propto (1+z)^{-1\sim-1.5}$ for other galaxy populations such as optical emission line galaxies (oELGs; e.g., \citealt{2017MNRAS.465.2717P}) and LBGs \citep[e.g., ][]{2004ApJ...600L.107F, 2012ApJ...756L..12M,2013ApJ...777..117M,2013ApJ...777..155O,2015ApJS..219...15S}. 

\begin{figure}[t!]
  \begin{center}
    \includegraphics[width=80mm]{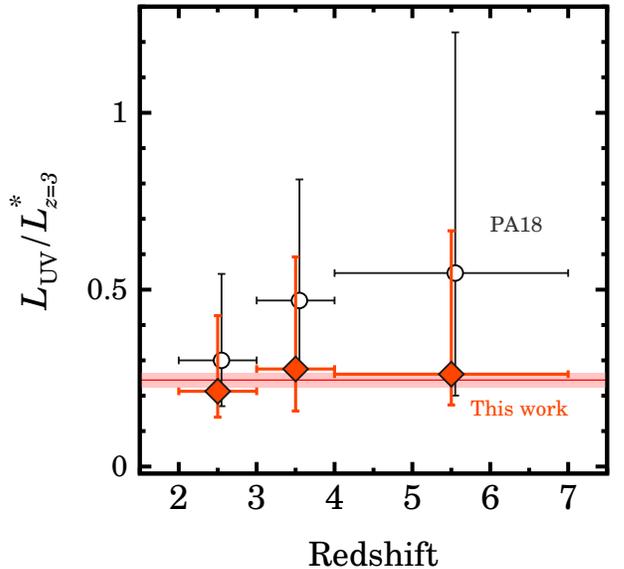}
  \end{center}
  \caption[]{{\footnotesize Median UV luminosity as a function of redshift for LAE samples of ours (red filled diamonds) and \citet{2018MNRAS.476.5479P}'s (open circles). The horizontal line and shaded region denotes the average $L_{\rm UV}$ value and its $1\sigma$ uncertainty of our median $L_{\rm UV}$ data points, respectively. The y-axis shows the UV luminosity in units of $L_{z=3}^*$. The error bars represent the 16th- and 84th-percentiles of the $L_{\rm UV}$ distribution.  }}
  \label{fig_z_luv}
\end{figure}

When considering the size evolution, it should be noted that the effective radius is strongly correlated with the galaxy luminosity (or galaxy mass) in the sense that a brighter source has a larger $r_{\rm e}$. Galaxy sizes decrease with redshift following this {\it size-luminosity relation} \citep[e.g., ][]{2014ApJ...788...28V, 2015ApJS..219...15S}. Contrary to expectations from the no size evolution of LAEs, several studies have reported that LAEs follow the evolving size-luminosity relation in samples of the Lyman alpha reference sample (LARS) project at $z\simeq0$ \citep{2015AA...576A..51G}, integral field unit (IFU) surveys at $z\simeq3-6$ \citep{2017AA...608A...8L,2016AA...587A..98W}, narrow-band (NB) observations at $z\simeq6-7$ \citep{2016ApJ...816...16J}. These results are incompatible with the lack of size evolution in LAEs. Thus, there is a possibility that the no size evolution results for LAEs are originated from a bias caused by heterogenous luminosity bins for a $r_{\rm e}$ comparison. To conclude whether the size evolution exists, we need to examine the size growth rate at {\it a given luminosity}. Recently, large LAE samples have been constructed by wide-field and deep surveys, as described in later sections. The combinations of the large LAE samples, the $\!${\it HST} deep data, and a systematic analysis will allow us to study the morphological evolution for LAEs with no significant bias.

\begin{deluxetable*}{cccccccccc}
\setlength{\tabcolsep}{0.35cm} 
\tabletypesize{\scriptsize}
\tablecaption{LAE Samples used for our Size Measurements}
\tablehead{\colhead{Reference} & \colhead{$z$} & \colhead{$N_{\rm LAE}$} & \colhead{$N_{\rm HST}^{\rm UV}$} & \colhead{$N_{\tt GALFIT}^{\rm UV}$} & \colhead{$N_{\rm HST}^{\rm opt}$} & \colhead{$N_{\tt GALFIT}^{\rm opt}$} & \colhead{$EW_{\rm 0,Ly\alpha}^{\rm limit}$} & \colhead{{\it HST} Field} & \colhead{Related Ref.}\\
\colhead{}& \colhead{}& \colhead{}& \colhead{} & \colhead{} & \colhead{} & \colhead{} & \colhead{[\AA]} & \colhead{} & \colhead{}\\
\colhead{(1)}& \colhead{(2)}& \colhead{(3)}& \colhead{(4)} & \colhead{(5)} & \colhead{(6)} & \colhead{(7)} & \colhead{(8)}&  \colhead{(9)} & \colhead{(10)}} 

\startdata 
\citet{2017ApJ...848..108W} & $0.3$ & $173$ & --- & --- & $18$ & $13$ & --- & cos, gds, aeg & Deh08, Cow10\\ 
\citet{2014ApJ...783..119W} & $1$ & $135$ & --- & --- & $9$ & $4$ & --- & cos, gds, aeg & Cow11, Bar12 \\
\citet{2012ApJ...745...12N} & $2.2$ & $3373$ & $591$ & $323$ & $503$ & $268$ & $>20-30$ & cos, uds, gds, gdn & \\
\citet{2012ApJ...744..110C} & $3.1$ & $199$ & $56$ & $32$ & --- & --- & $>20$ & gds & Gro07 \\
\citet{2017arXiv171204451S} & $3.1$ & $45$ & $8$ & $4$ & --- & --- & $>25$ & cos & \\
                                                & $2.4-2.6^{\rm b}$ & $741$ & $43$ & $30$ & --- & --- & $>50$ & cos & \\
                                                & $2.7-2.9^{\rm b}$ & $311$ & $9$ & $8$ & --- & --- & $>50$ & cos & \\
                                                & $2.9-3.1^{\rm b}$ & $711$ & $41$ & $21$ & --- & --- & $>50$ & cos & \\
                                                & $3.1-3.3^{\rm b}$ & $483$ & $26$ & $17$ & --- & --- & $>50$ & cos & \\
                                                & $3.2-3.4^{\rm b}$ & $641$ & $26$ & $8$ & --- & --- & $>50$ & cos & \\
                                                & $3.6-3.9^{\rm b}$ & $98$ & $2$ & $0$ & --- & --- & $>50$ & cos & \\
                                                & $4.0-4.3^{\rm b}$ & $142$ & $6$ & $0$ & --- & --- & $>50$ & cos & \\
                                                & $4.4-4.7^{\rm b}$ & $79$ & $4$ & $2$ & --- & --- & $>50$ & cos & \\
                                                & $4.7-5.0^{\rm b}$ & $81$ & $6$ & $3$ & --- & --- & $>50$ & cos & \\
                                                & $4.9-5.2^{\rm b}$ & $79$ & $3$ & $1$ & --- & --- & $>50$ & cos & \\
                                                & $5.2-5.5^{\rm b}$ & $33$ & $1$ & $0$ & --- & --- & $>50$ & cos & \\
                                                & $5.6-5.9^{\rm b}$ & $35$ & $0$ & $0$ & --- & --- & $>50$ & cos & \\
\citet{2009ApJ...691..465F} & $4.5$ & $14$ & $14$ & $7$ & --- & --- & $\gtrsim50$ & gds & \\
\citet{2009ApJ...700..899S} & $4.9$ & $79$ & $2$ & $1$ & --- & --- & $>11$ & cos & \\
\citet{2008ApJS..176..301O} & $3.1$ & $356$ & $15$ & $1$ & --- & --- & $\gtrsim64$ & uds & \\
                                              & $3.7$ & $101$ & $7$ & $2$ & --- & --- & $\gtrsim44$ & uds & \\
                                              & $5.7$ & $401$ & $18$ & $5$ & --- & --- & $\gtrsim27$ & uds & \\
\citet{2007ApJS..172..523M} & $5.7$ & $119$ & $4$ & $0$ & --- & --- & $>18$ & cos & \\
\citet{2010ApJ...725..394H} & $5.7$ & $88$ & $4$ & $0$ & --- & --- & --- & gdn & \\
                                             & $6.5$ & $30$ & $0$ & $0$ & --- & --- & --- & gdn & \\
\citet{2010ApJ...723..869O} & $6.6$ & $207$ & $14$ & $5$ & --- & --- & $\gtrsim14$ & uds & \\
\citet{2015MNRAS.451..400M} & $6.6$ & $16$ & $2$ & $1$ & --- & --- & $>38$ & cos & \\
\citet{2017AA...608A...8L}$^{\rm a}$ & $2.9-6.6$ & $184$ & $180$ & $97$ & --- & --- & $\gtrsim20$ & hudf & \\
\citet{2018MNRAS.473...30C}$^{\rm a}$ & $2.9-6.6$ & $100$ & $100$ & $78$ & --- & --- & --- & gds & Her17 \\
\citet{2018arXiv180505944I} & $7$ & $34$ & $4$ & $1$ & --- & --- & $\gtrsim10$ & cos, uds & Ino \\
\citet{2017ApJ...844...85O} & $7$ & $20$ & $2$ & $1$$^{\rm c}$ & --- & --- & $>10$ & uds & \\
\citet{2017ApJ...845L..16H} & $7$ & $6$ & $0$ & $0$ & --- & --- & $>10$ & cos & Zhe17 \\
\citet{2018arXiv180801847P}$^{\rm d}$ & $7$ & $2$ & $2$ & $1$ & --- & --- & --- & cos, uds, gds & \\
\citet{2012ApJ...752..114S} & $7.2$ & $1$ & $0$ & $0$ & --- & --- & --- & uds & \\
\citet{2012ApJ...744...83O} & $7.2$ & $1$ & $1$ & $1$ & --- & --- & --- & gdn & Ouc09 \\
\citet{2015ApJ...804L..30O} & $7.7$ & $1$ & $1$ & $1$ & --- & --- & --- & aeg & Smi15 \\
\hline 
Total & --- & $9119$ & $1192$ & $651$ & $530$ & $285$ & --- & --- & ---
\enddata

\tablecomments{Columns: (1) Reference. (2) Redshift range of the LAE sample. (3) Number of LAEs in the sample. (4) Number of LAEs whose rest-frame UV continuum emission is covered by a passband of the {\it HST} images. (5) Number of LAEs whose $r_{\rm e}^{\rm UV}$ value is obtained in our size measurements. (6) Number of LAEs whose rest-frame optical continuum emission is covered by a passband of the {\it HST} images. (7) Number of LAEs whose $r_{\rm e}^{\rm Opt}$ value is obtained in our size measurements. (8) Typical Ly$\alpha$ EW limit of the LAE survey. (9) {\it HST} field (``gds": GOODS-South; ``gdn": GOODS-North; ``uds": UDS; ``aeg": AEGIS; ``cos": COSMOS; ``hudf": Hubble Ultra Deep Field). (10) Reference related to the LAE sample (``Deh08": \citealt{2008ApJ...680.1072D}; ``Cow10": \citealt{2010ApJ...711..928C}; ``Cow11": \citealt{2011ApJ...735L..38C}; ``Bar12": \citealt{2012ApJ...749..106B}; ``Gro07": \citealt{2007ApJ...667...79G}; ``Her17": \citealt{2017A&A...606A..12H}; ``Ino": A. K. Inoue et al. in preparation: ``Zhe17": \citealt{2017ApJ...842L..22Z}; ``Ouc09": \citealt{2009ApJ...706.1136O}; ``Smi15": \citealt{2015ApJ...801..122S})}
\tablenotetext{a}{IFU observations.}
\tablenotetext{b}{MB filter observations.}
\tablenotetext{c}{This object is identical to a $z\simeq7$ LAE selected by \citet{2018arXiv180505944I}. }
\tablenotetext{d}{Only the galaxies at $z>7$ are selected. }
\label{tab_lae_sample}
\end{deluxetable*}

\begin{figure}[t!]
  \begin{center}
    \includegraphics[width=85mm]{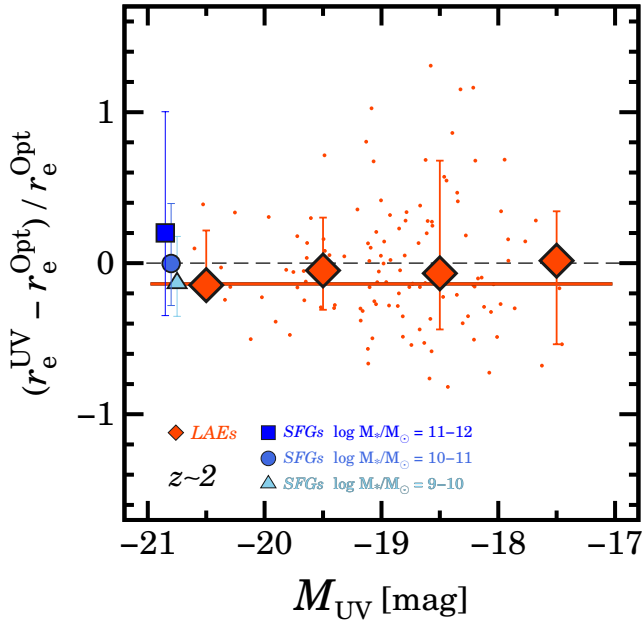}
  \end{center}
  \caption[]{{\footnotesize Difference between $r_{\rm e}^{\rm UV}$ and $r_{\rm e}^{\rm Opt}$ for the LAEs at $z\simeq2.2$ (red dots) as a function of UV magnitude. The red diamonds with error bars represent the median values of $(r_{\rm e}^{\rm UV} - r_{\rm e}^{\rm Opt})/r_{\rm e}^{\rm Opt}$ in different $M_{\rm UV}$ bins. The horizontal line shows the average value of median $(r_{\rm e}^{\rm UV} - r_{\rm e}^{\rm Opt})/r_{\rm e}^{\rm Opt}$ data points. The blue symbols represent median values of $(r_{\rm e}^{\rm UV} - r_{\rm e}^{\rm Opt})/r_{\rm e}^{\rm Opt}$ for $z\simeq2$ SFGs with $M_{\rm UV}=-21-(-20)$ in different stellar mass ranges (triangle: $\log M_* / M_\odot = 9-10$; circle: $\log M_* / M_\odot = 10-11$; square: $\log M_* / M_\odot = 11-12$; see \citetalias{2015ApJS..219...15S}). The data points are slightly shfited along the $x$-axis for clarity. }}
  \label{fig_muv_ruvropt_lae}
\end{figure}

In this paper, we investigate the redshift evolution of the radial SB profile and the galaxy size with a sample of $\simeq9,000$ LAEs at $z\simeq0-8$ and the $\!${\it HST} deep data of extra-galactic legacy surveys. This is the third paper in a series studying the morphology of high-$z$ galaxies.\footnote{The first and second papers examine galaxy sizes \citep[][ hereafter \citetalias{2015ApJS..219...15S}]{2015ApJS..219...15S} and clumpy structures \citep[][ hereafter \citetalias{2016ApJ...821...72S}]{2016ApJ...821...72S} using $\simeq190,000$ galaxies at $z\simeq0-10$, respectively. } This paper is organized as follows. In Sections \ref{sec_sample} and \ref{sec_data}, we describe the details of our LAE sample and {\it HST} data, respectively. Section \ref{sec_analysis} presents methods to obtain radial SB profiles and $r_{\rm e}$. In Section \ref{sec_comparison}, we detail comparison samples of UV selected galaxies and LAEs in the literature. We present the redshift evolution of the radial SB profiles, the size-luminosity relation, $r_{\rm e}$, and size-relevant physical quantities in Section \ref{sec_result}. Section \ref{sec_discuss} discusses the implications for the galaxy formation and evolution. In Section \ref{sec_conclusion}, we summarize the findings in our study. 

Throughout this paper, we adopt the concordance cosmology with $(\Omega_m, \Omega_\Lambda, h)=(0.3, 0.7, 0.7)$, \citep{2016AA...594A..13P}. All magnitudes are given in the AB system \citep{1983ApJ...266..713O}. We refer to the {\it HST} F606W, F775W, F814W, F850LP, F098M, F105W, F125W, F140W, and F160W filters as $V_{606}, I_{814}, z_{850}, Y_{098}, Y_{105}, J_{125}, JH_{140}$, and $H_{160}$, respectively. 

\begin{figure}[t!]
  \begin{center}
    \includegraphics[width=85mm]{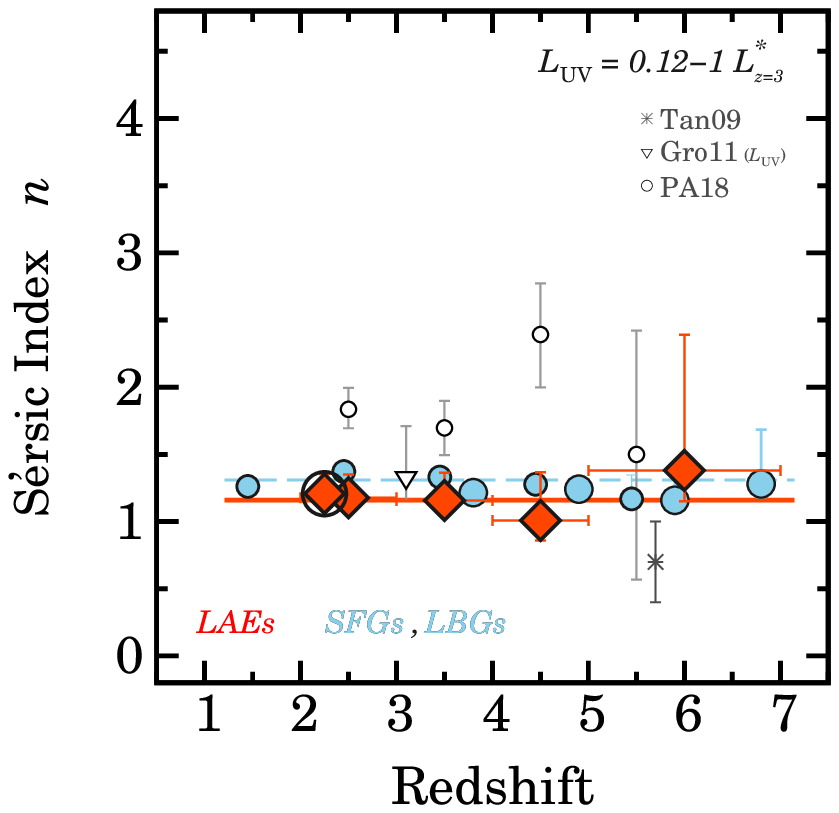}
  \end{center}
  \caption[]{{\footnotesize S\'ersic index as a function of redshift in the UV luminosity bin of $L_{\rm UV}=0.12-1 L_{z=3}^*$. The red filled diamonds with and without an open circle represent the S\'ersic index for LAEs in the rest-frame UV and optical continuum emission, respectively. The gray symbols indicate LAEs in previous studies (gray asterisks: \citealt{2009ApJ...701..915T}; gray open circles: \citealt{2018MNRAS.476.5479P}; gray open inverse triangles: \citealt{2011ApJ...743....9G}). The measurement technique is noted in the parenthesis of the legend (S: {\tt SExtractor}; G: {\tt GALFIT}; see also Table \ref{tab_previous_studies}). The median $n$ value is calculated in the bin of $L_{\rm UV}=0.12-1 L_{z=3}^*$ for \citet{2011ApJ...743....9G} with $n$ for individual sources, which is indicated as ``$L_{\rm UV}$" in the parenthesis of the legend. The small and large cyan filled circles denote the SFGs and LBGs, respectively (\citetalias{2015ApJS..219...15S}; see also Section \ref{sec_result_rad}). The magenta solid and cyan dashed horizontal line denote weighted means of $\left< n\right>=1.16$ for the LAEs and $\left< n\right>=1.31$ for the SFGs and LBGs, respectively. Our LAEs are not plotted at $z\lesssim2$, because the number of LAEs with $L_{\rm UV}=0.12-1 L_{z=3}^*$ is too small. The data points are slightly shifted along the $x$-axis for clarity. The error bars of some data points are smaller than the size of symbols.}}
  \label{fig_z_n_lae}
\end{figure}

\begin{figure*}[t!]
  \begin{center}
    \includegraphics[width=182mm]{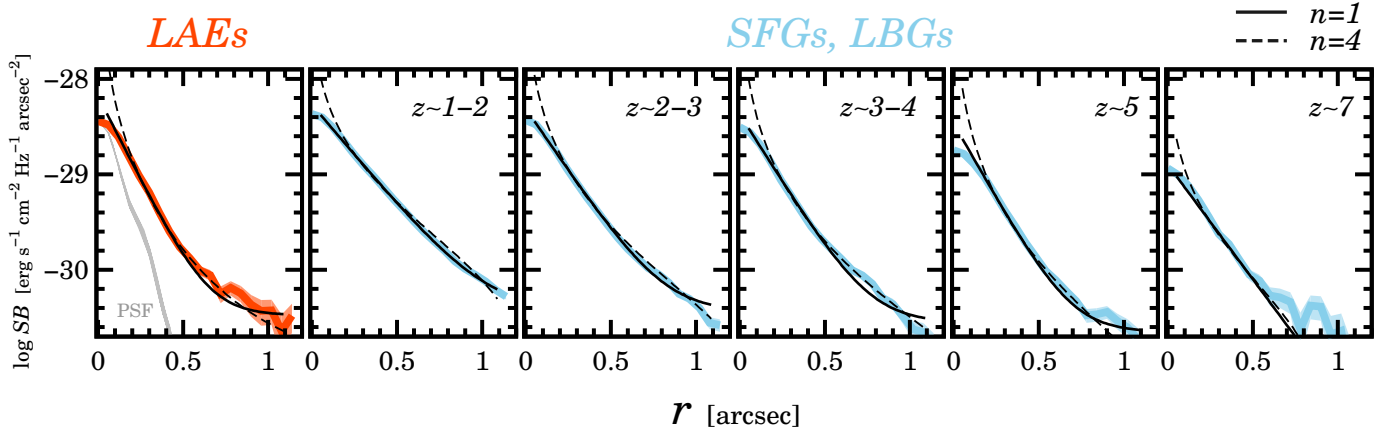}
  \end{center}
  \caption[]{{\footnotesize Radial SB profiles at $\lambda_{\rm UV}$ for LAEs (red), SFGs, and LBGs (cyan) in the UV luminosity bin of $L_{\rm UV}=0.12-1 L_{z=3}^*$. The panels, from left to right, indicate the LAEs at $z\simeq0-7$, SFGs at $z=1-2$,  $z=2-3$, $z=3-4$, LBGs at $z\simeq5$, and $z\simeq7$. The shaded regions show the 1$\sigma$ uncertainty of the radial SB profiles. The solid and dashed black curves depict the best-fit S\'ersic profiles with $n=1$ and $n=4$, respectively. The gray line shows a typical PSF of the $H_{160}$ image. }}
  \label{fig_rad_prof_all}
\end{figure*}

\section{LAE Sample}\label{sec_sample}

To perform a statistical study on galaxy structures, we construct a large LAE sample by combining several catalogs of LAEs at $z\simeq0-8$ in the literature: LAEs obtained from {\it GALEX} space-based spectroscopy (\citealt{2017ApJ...848..108W, 2014ApJ...783..119W}), ground-based NB and medium-band (MB) imaging (\citealt{2012ApJ...745...12N, 2012ApJ...744..110C, 2017arXiv171204451S, 2009ApJ...691..465F, 2009ApJ...700..899S, 2008ApJS..176..301O, 2007ApJS..172..523M, 2010ApJ...725..394H, 2010ApJ...723..869O, 2015MNRAS.451..400M, 2018arXiv180505944I, 2017ApJ...844...85O}), IFU observations with the Multi Unit Spectroscopic Explorer \citep[MUSE; ][]{2010SPIE.7735E..08B} on the Very Large Telescope (\citealt{2017AA...608A...8L, 2018MNRAS.473...30C}), and spectroscopic observations for $z\simeq7-8$ \citep{2017ApJ...845L..16H, 2018arXiv180801847P, 2012ApJ...744...83O, 2012ApJ...752..114S, 2015ApJ...804L..30O}. The $z\simeq7-8$ LAEs are not used, but for a $z$-$r_{\rm e}$ diagram, because of the small statistics. The sample contains $9119$ LAEs in total. Table \ref{tab_lae_sample} lists the redshift range and the object number of each LAE catalog. Figure \ref{fig_muv_ewlya} shows the absolute UV magnitude, $M_{\rm UV}$, and the rest-frame Ly$\alpha$ equivalent width (EW), $EW_{\rm 0, Ly\alpha}$. Thanks to the combination of space-based spectroscopic, ground-based imaging and IFU data, the LAE sample covers wide ranges of $M_{\rm UV}$ and $EW_{\rm 0, Ly\alpha}$. 

We find some features of the heterogeneity in $M_{\rm UV}$ and $EW_{\rm 0,Ly\alpha}$ between low-$z$ and high-$z$ samples. For example, as clearly shown at $M_{\rm UV}\simeq-18-(-16)$ in Figure \ref{fig_muv_ewlya}, faint sources tend to be preferentially selected at low-$z$. See also Figure \ref{fig_muv_ewlya_z} which is the same as Figure \ref{fig_muv_ewlya}, but for  color-coded symbols based on the redshift. To avoid this selection effect, we apply a UV luminosity bin of $0.12-1$ $L_{\rm UV}/L_{z=3}^*$, where $L_{z=3}^*$ is the characteristic UV luminosity of LBGs at $z\sim3$ \citep[$M_{\rm UV}=-21$, ][]{1999ApJ...519....1S}. To exclude several objects with a relatively low Ly$\alpha$ EW, we also select LAEs with $EW_{\rm 0,Ly\alpha}>20$\AA\, which is the well-used $EW_{\rm 0,Ly\alpha}$ criterion for LAE surveys. 

In Figure \ref{fig_z_luv}, we check whether there is no significant difference in typical $L_{\rm UV}$ values in the redshift range. Our LAE sample shows a nearly constant $L_{\rm UV}$ value of $L_{\rm UV}/L_{z=3}^*\simeq0.25$. We also plot the median $L_{\rm UV}$ of \citet{2018MNRAS.476.5479P} which is the most recent statistical study on the LAE size evolution. In contrast to our LAE sample, \citet{2018MNRAS.476.5479P}'s $L_{\rm UV}$ values increase from $L_{\rm UV}/L_{z=3}^*\simeq0.25$ at $z=2-3$ to $L_{\rm UV}/L_{z=3}^*\simeq0.5$ at $z=4-7$. Section \ref{sec_discuss_size_evolution} discusses the importance of the $L_{\rm UV}$ cut on the galaxy size evolution. 

\begin{deluxetable*}{cccccc}
\setlength{\tabcolsep}{0.35cm} 
\tabletypesize{\scriptsize}
\tablecaption{Statistical Studies on LAE sizes in the Literature}
\tablehead{\colhead{Reference} & \colhead{$N_{\tt GALFIT} (N_{\rm LAE})$} & \colhead{$z$} & \colhead{Size Measurements} & \colhead{LAE Catalog}\\
\colhead{(1)}& \colhead{(2)}& \colhead{(3)}& \colhead{(4)} & \colhead{(5)}} 

\startdata
\citet{2015AA...576A..51G} & $14$ $(14)$ & $0.1$ & {\tt SExtractor}, {\tt PHOT}, {\tt ELLIPSE} & \cite{2014ApJ...797...11O}\\
\citet{2017ApJ...838....4Y} & $24$ $(43)$ & $0.1-0.3$ & FWHM & \citet{2017ApJ...844..171Y}\\
\citet{2016ApJ...817...79H} & $28$ $(28)$ & $2$ & {\tt PHOT} & compilation\\
\citet{2014ApJ...785...64S} & $663$ $(1239)$ & $2.2$ & {\tt SExtractor} & \citet{2012ApJ...745...12N}\\
\citet{2009ApJ...705..639B} & \nodata $(120)$ & $3.1$ & {\tt PHOT} & \citet{2007ApJ...667...79G}\\
\citet{2011ApJ...743....9G} & $78$ $(78)$ & $3.1$ & {\tt GALFIT} & \citet{2007ApJ...667...79G}\\
\citet{2014ApJ...786...59H} & $63$ $(99)$ & $1.9-3.6$ & {\tt PHOT} & \citet{2011ApJ...736...31B}\\
\citet{2017MNRAS.468.1123S} & $50$ $(50)$ & $2.5$ & {\tt GALAPAGOS} & \citet{2017MNRAS.468.1123S}\\
\citet{2012ApJ...753...95B} & \nodata $(108)$ & $2.1$ & {\tt PHOT} & \citet{2007ApJ...667...79G}\\
                                           & \nodata $(171)$ & $3.1$ & {\tt PHOT} & \citet{2012ApJ...744..110C}\\
\citet{2016ApJ...819...25K} & 54 $(61)$ & $4.9$ & {\tt SExtractor} & \citet{2009ApJ...700..899S}\\
\citet{2009ApJ...701..915T} & $47$ $(119)$ & $5.7$ & {\tt SExtractor} & \cite{2007ApJS..172..523M}\\
\citet{2012ApJ...750L..36M} & 174 $(174)$ & $2.35-6$ & compilation & compilation\\
\citet{2007ApJ...667...49P} & \nodata $(9)$ & $4-5.7$ & {\tt GALFIT} & \citet{2007AJ....134..169X}\\
\citet{2013ApJ...773..153J} & \nodata $(51)$ & $5.7-7$ & {\tt SExtractor} & compilation\\
\citet{2016AA...587A..98W} & \nodata $(26)$ & $2.9-6.6$ & exponential fit & \citet{2015AA...575A..75B}\\
\citet{2017AA...608A...8L} & \nodata $(145)$ & $2.9-6.6$ & exponential fit & \citet{2017AA...608A...1B}\\
\citet{2018MNRAS.476.5479P} & $429$ $(3045)$ & $2-6$ & {\tt GALFIT} & compilation\\
This work & 936 $(9119)$ & $0.3-7.7$ & {\tt GALFIT} & compilation
\enddata

\tablecomments{Columns: (1) Reference. (2) Number of galaxies whose effective radius is measured in the reference. The values in parentheses are the number of galaxies in the parent sample. (3) Redshift range for size measurements of LAEs. (4) Method or software to measure galaxy sizes. See the text in Section \ref{sec_comparison}. (5) Reference of LAE catalog used for the size measurements. The ``compilation" indicates that the study uses an LAE sample compiled from several LAE catalogs. }
\label{tab_previous_studies}
\end{deluxetable*}

\section{HST Data}\label{sec_data}

We make use of deep {\it HST} imaging data taken by the Cosmic Assembly Near-infrared Deep Extragalactic Legacy Survey \citep[CANDELS; ][]{2011ApJS..197...35G, 2011ApJS..197...36K}. The CANDELS consists of five deep fields, GOODS-South, GOODS-North, UDS, AEGIS, and COSMOS. These $\!${\it HST} images are retrieved from the website of the 3D-HST project \citep{2014ApJS..214...24S}.\footnote{http://3dhst.research.yale.edu/Home.html} We also use the {\it Hubble} Ultra Deep Field 09+12 \citep[HUDF 09+12; ][]{2006AJ....132.1729B, 2011ApJ...737...90B, 2013ApJS..209....6I, 2013ApJ...763L...7E}\footnote{http://archive.stsci.edu/prepds/xdf/} to examine LAEs in the MUSE deep field \citep{2017AA...608A...8L}. 

The typical $5\sigma$ limiting magnitudes in a $0.\!\!^{\prime\prime}35$ diameter aperture are $\simeq28-29$ mag for CANDELS and $\simeq30$ mag for HUDF 09$+$12. The full width half maximum (FWHM) of the point spread function (PSF) are $\simeq0.\!\!^{\prime\prime}08-0.\!\!^{\prime\prime}09$ and $\simeq0.\!\!^{\prime\prime}12-0.\!\!^{\prime\prime}18$ in the Advanced Camera for Surveys (ACS) and Wide Fields Camera 3 (WFC3)/IR images, respectively. The full set of the limiting magnitudes and PSF FWHM is provided in Table 1 of \citetalias{2015ApJS..219...15S}. 

\section{Analysis}\label{sec_analysis}

Using the $\!${\it HST} images, we measure the structural quantities for individual sources of the LAE sample. The method of the analysis is the same as that in \citetalias{2015ApJS..219...15S}, but is briefly described here. First, we extract $18^{\prime\prime}\times18^{\prime\prime}$ cutout images from the {\it HST} data at the position of each LAE. Next, we obtain structural quantities such the half-light radius along the semi-major axis, $R_{\rm e,major}$, the S\'ersic index, $n$, and the position angle (P.A.), the axial ratio, $q$, by performing the two-dimensional (2D) S\'ersic profile \citep{1963BAAA....6...41S,1968adga.book.....S} fitting with the {\tt GALFIT} software \citep{2002AJ....124..266P,2010AJ....139.2097P}. The $R_{\rm e, major}$ is converted to the ``circularized" radius, $r_{\rm e}$, through $r_{\rm e}\equiv R_{\rm e, major}\sqrt{q}$. In the {\tt GALFIT} fitting, we input the {\it sigma}, {\it mask}, and PSF images. The {\it sigma} and {\it mask} images are used for the fitting weight of individual pixels and masking neighboring objects of the main galaxy components, respectively. The mask images also remove the light of galaxy clumpy structures, enabling us to focus on analyses for the main galaxy components. The PSF images of each $\!${\it HST} field are provided in the 3D-HST project \citep{2014ApJS..214...24S}. The ranges of the structural parameters varying in the fitting are $\Delta m<3$ mag, $0.3<R_{\rm e, major}<400$ pixels, $0.2<n<8$, $0.0001<q<1$, $\Delta x<4$ pixel, and $\Delta y<4$ pixel, where $m$ and $(x, y)$ are the magnitude and the coordinates in the {\it HST} images. The objects are discarded if one or more fitting parameters reach the limit of the parameter ranges (e.g., $R_{\rm e, major}=400$ pixel). The cosmological SB dimming would not significantly affect  our measurements of structural quantities, at least at $M_{\rm UV}\lesssim-18$, which has been evaluated in \citetalias{2015ApJS..219...15S}. To avoid a possible profile fitting degeneracy between $r_{\rm e}$ and $m$ obtained in {\tt GALFIT}, we employ {\tt MAG\_AUTO} derived with SExtractor as a total magnitude. See \citetalias{2015ApJS..219...15S} for more details of our analysis. \footnote{Some recent studies have performed a sophisticated technique to simultaneously determine the size-luminosity relation and the UV luminosity functions \citep[e.g., ][]{2018ApJ...855....4K}. }

To minimize the effect of morphological {\it K}-correction for LAEs at different redshifts of $z\simeq0-8$, we utilize several bands of images taken with ACS and WFC3/IR on {\it HST}. Based on the redshift of LAEs, we select a passband from $V_{606}$, $I_{814}$, $J_{125}$, $H_{160}$, or the coadd WFC3 band for covering the wavelength ranges of the rest-frame UV, $\lambda_{\rm UV}\simeq1500-3000$\,\AA\,, or the rest-frame optical, $\lambda_{\rm Opt}\simeq4500-8000$\,\AA, emission. The effective radius measured in $\lambda_{\rm UV}$ and $\lambda_{\rm Opt}$ are referred to as $r_{\rm e}^{\rm UV}$ and $r_{\rm e}^{\rm Opt}$, respectively. In {\it HST} fields where the $Y_{105}$ band is available (i.e., HUDF 09$+$12, GOODS-South, and GOODS-North), we use coadded images constructed from four or five WFC3/IR bands \citep{2016ApJ...821..123H}. 

Table \ref{tab_lae_sample} summarizes the numbers of LAEs that are covered by the {\it HST} images and are well fitted by {\tt GALFIT}. The total numbers of LAEs whose structural quantities are obtained in $\lambda_{\rm UV}$ and $\lambda_{\rm Opt}$ are 651 and 285, respectively. We have excluded candidates of active galactic nuclei by checking X-ray and radio wavelength source catalogs of e.g., the {\it XMM-Newton} $0.2-10.0$ keV band and Very Large Array 1.4 GHz (see \citealt{2012ApJ...745...12N, 2016ApJ...823...20K} for more details). 

Here we compare $r_{\rm e}^{\rm UV}$ and $r_{\rm e}^{\rm Opt}$ of the LAEs at $z\simeq2.2$ where both the radii can be obtained with the $\!${\it HST} data. Figure \ref{fig_muv_ruvropt_lae} shows the differences between $r_{\rm e}^{\rm UV}$ and $r_{\rm e}^{\rm Opt}$ as a function of $M_{\rm UV}$. Albeit with a large scatter, the median values of $(r_{\rm e}^{\rm UV}-r_{\rm e}^{\rm Opt})/r_{\rm e}^{\rm Opt}$ are $\lesssim20$ \% in all the $M_{\rm UV}$ bins. This indicates that the effect of morphological {\it K}-correction are small for LAEs at $z\simeq2.2$. Although $r_{\rm e}^{\rm UV}$ agrees well with $r_{\rm e}^{\rm Opt}$, the two radii are analyzed independently for measurements of e.g., the size growth rate, in the following sections. The $r_{\rm e}^{\rm Opt}$ data points are used just for a reference. 

For comparison, we plot  $z\simeq2$ SFGs of \citetalias{2015ApJS..219...15S} in Figure \ref{fig_muv_ruvropt_lae}. In \citetalias{2015ApJS..219...15S}, we have found that $r_{\rm e}^{\rm UV}$ relative to $r_{\rm e}^{\rm Opt}$ is larger for more massive SFGs in a stellar mass range of $\log{M_*/M_\sun}\simeq9-12$. This trend could be interpreted as large dust attenuation in the galactic central regions \citep[e.g., ][]{2012MNRAS.421.1007K} and/or the inside-out disk formation \citep[e.g., ][]{2009ApJ...697.1290B, 2009ApJ...694..396B} for massive SFGs. Similar to the low mass SFGs with $\log{M_*/M_\sun}\simeq9-10$, our LAEs show, on average, a small $r_{\rm e}^{\rm UV}$ relative to $r_{\rm e}^{\rm Opt}$. According to an $M_{\rm UV}$-$M_*$ empirical relation in \citetalias{2015ApJS..219...15S}, the stellar mass for our LAEs corresponds to $\log{M_*/M_\odot} \simeq 9-10$ that is comparable to $M_*$ for the low mass SFGs. Thus, the consistency in $r_{\rm e}^{\rm UV}$/$r_{\rm e}^{\rm Opt}$ and $M_*$ might suggest that LAEs typically have small amounts of dust at the galactic central regions and/or have not experienced the inside-out disk formation. 

\begin{figure}[t!]
  \begin{center}
    \includegraphics[width=85mm]{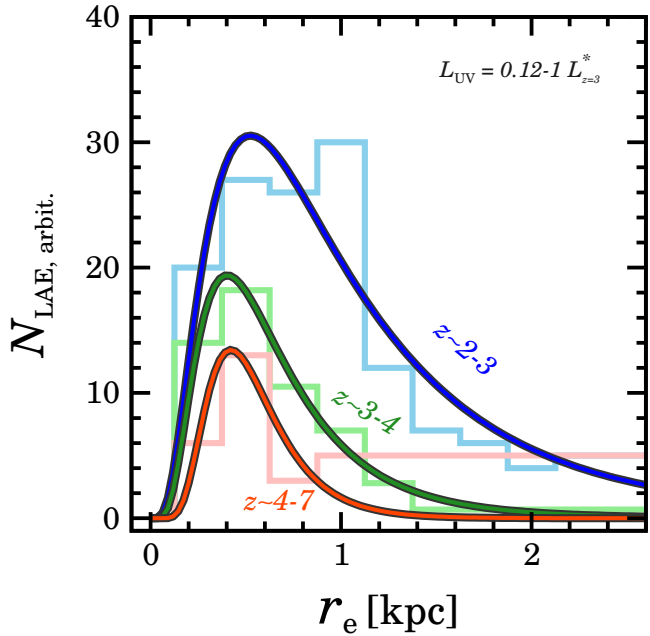}
  \end{center}
  \caption[]{{\footnotesize Distribution of $r_{\rm e}^{\rm UV}$ for the LAEs in the UV luminosity bin of $L_{\rm UV}=0.12-1\,L_{z=3}^*$. The histograms and the curves show the $r_e^{\rm UV}$ distributions and the best-fit log-normal functions, respectively, for the LAEs at $z\simeq2-3$ (blue), $z\simeq3-4$ (green), and $z\simeq4-7$ (red). The y-axis is arbitrary. }}
  \label{fig_hist_re_L012_1_lae_all}
\end{figure}

\begin{figure*}[t!]
  \begin{center}
    \includegraphics[width=180mm]{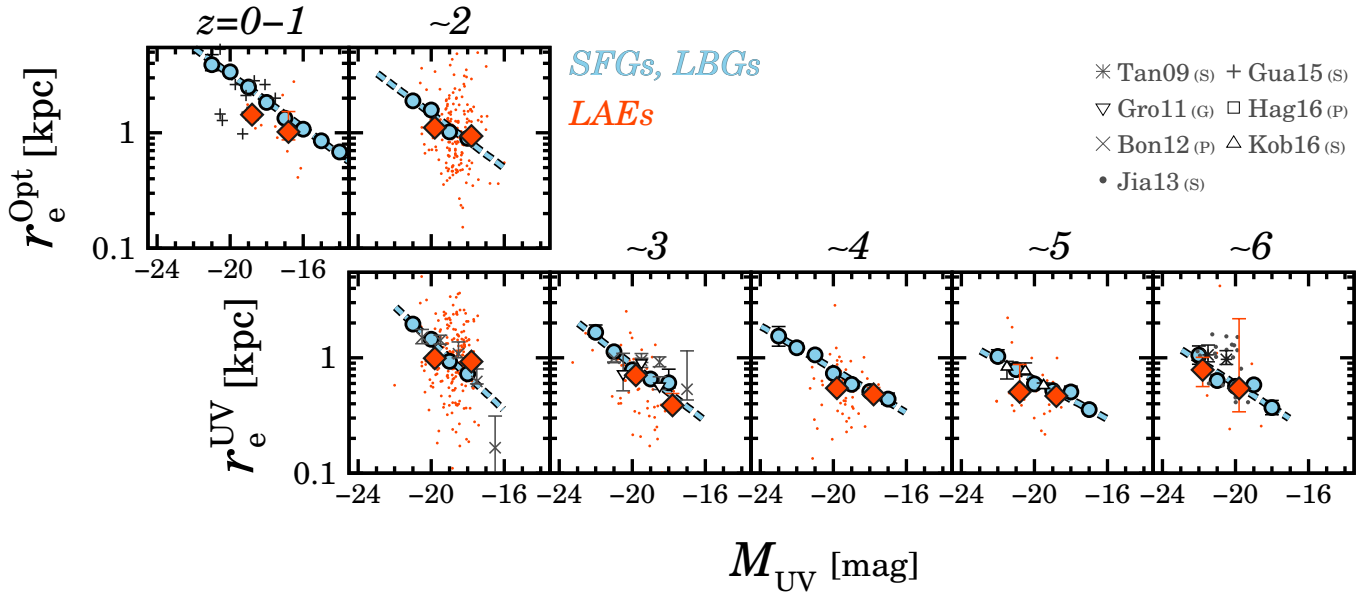}
  \end{center}
  \caption[]{{\footnotesize Effective radius, $r_{\rm e}$ and UV magnitude $M_{\rm UV}$ relation. The top and bottom panels represent $r_{\rm e}^{\rm Opt}$ and $r_{\rm e}^{\rm UV}$, respectively. The redshifts are labeled at the top of the panels. The red filled diamonds and dots indicate the representative and individual $r_{\rm e}$ measurements for the LAEs. The cyan filled circles represent the SFGs and LBGs (\citetalias{2015ApJS..219...15S}). The cyan dashed lines denote the best-fit power-law functions of $r_{\rm e}\propto L_{\rm UV}\,^\alpha$ for the $r_{\rm e}$-$M_{\rm UV}$ relations. The gray symbols present LAEs in the literature (gray asterisks: \citealt{2009ApJ...701..915T}; gray open inverse triangles: \citealt{2011ApJ...743....9G}; gray x-marks: \citealt{2012ApJ...753...95B}; gray dots: \citealt{2013ApJ...773..153J}; gray crosses: \citealt{2015AA...576A..51G}; gray open squares: \citealt{2016ApJ...817...79H}; gray open triangles: \citealt{2016ApJ...819...25K}). The measurement technique is noted in the parenthesis of the legend (S: {\tt SExtractor}; G: {\tt GALFIT}; P: {\tt PHOT}; see also Table \ref{tab_previous_studies}). The data points are slightly shifted along the $x$-axis for clarity. The error bars of some data points are smaller than the size of symbols.}}
  \label{fig_muv_re_lae_all}
\end{figure*}

\begin{figure*}[t!]
  \begin{center}
    \includegraphics[width=165mm]{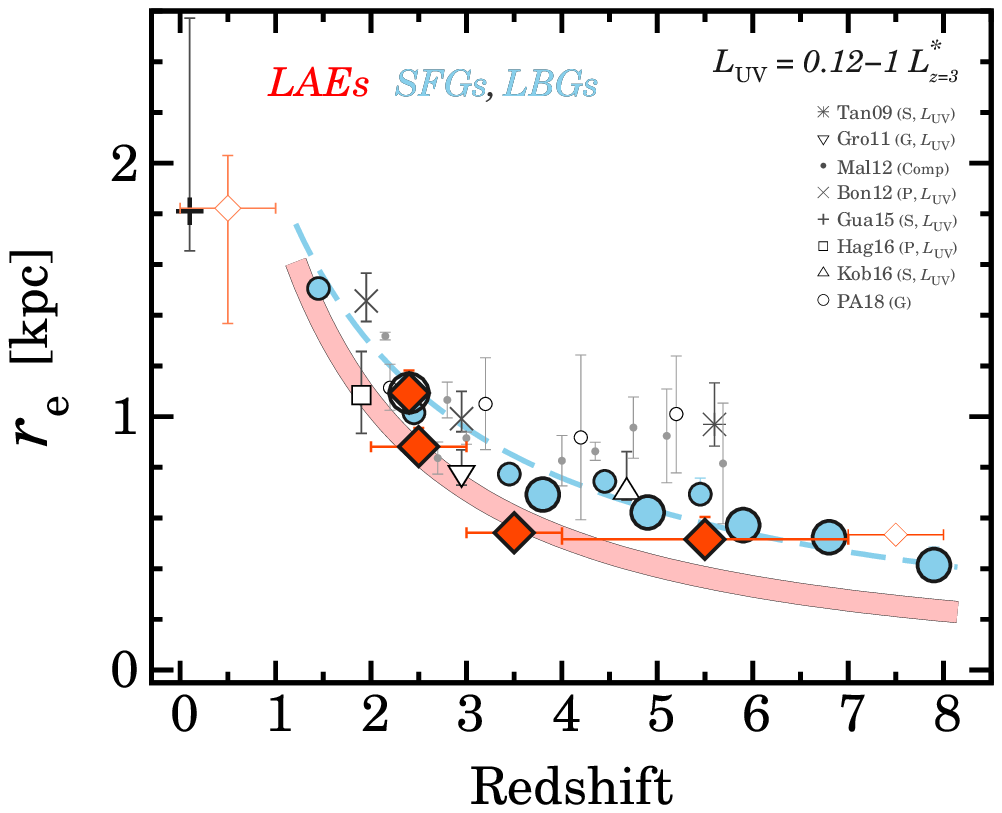}
  \end{center}
  \caption[]{{\footnotesize Redshift evolution of $r_{\rm e}$ in the $L_{\rm UV}$ range of $0.12-1$ $L_{z=3}^*$. The red filled diamonds with and without an open circle represent $r_{\rm e}^{\rm UV}$ and $r_{\rm e}^{\rm Opt}$ values measured for our LAEs, respectively. The red open diamond at $z\simeq0.5$ is $r_{\rm e}^{\rm Opt}$ for LAEs at $z\simeq0-1$ which is inferred from the extrapolation of the size-luminosity relation in Figure \ref{fig_muv_re_lae_all} (see also Section \ref{sec_result_muv_re} in more details). The red open diamond at $z\simeq7.5$ represents $r_{\rm e}^{\rm UV}$ measured from the three LAEs at $z\simeq7-8$. The error bars of $r_{\rm e}^{\rm UV}$ at $z\simeq7-8$ are not reliably estimated because of the small statistics. The small and large cyan filled circles indicate the SFGs and LBGs, respectively (\citetalias{2015ApJS..219...15S}). The magenta solid and cyan dashed lines present the best-fit $(1+z)^\beta$ functions for the LAEs and SFGs/LBGs, respectively. The best-fit $\beta$ value for $r_{\rm e}^{\rm UV}$ of the LAEs is $-1.37\pm0.65$, which is obtained from the three $r_{\rm e}^{\rm UV}$ data points at $z\simeq2-7$. The gray symbols present LAEs in the literature (gray asterisks: \citealt{2009ApJ...701..915T}; gray open inverse triangles: \citealt{2011ApJ...743....9G}; gray x-marks: \citealt{2012ApJ...753...95B}; gray dots: \citealt{2012ApJ...750L..36M}; gray crosses: \citealt{2015AA...576A..51G}; gray open squares: \citealt{2016ApJ...817...79H}; gray open triangles: \citealt{2016ApJ...819...25K}; gray open circles: \citealt{2018MNRAS.476.5479P}). The measurement technique is noted in the parenthesis of the legend (S: {\tt SExtractor}; G: {\tt GALFIT}; P: {\tt PHOT}; see also Table \ref{tab_previous_studies}). \citet{2012ApJ...750L..36M} compile several results of $r_{\rm e}$ measurements in the literature. For the previous studies with ``$L_{\rm UV}$" in the parenthesis of the legend, the median $r_{\rm e}$ value is calculated in the range of $L_{\rm UV}=0.12-1 L_{z=3}^*$. The data points are slightly shifted along the $x$-axis for clarity. The error bars of some data points are smaller than the size of symbols.}}
  \label{fig_z_re_lae}
\end{figure*}

\section{Comparison Samples}\label{sec_comparison}

To make a comparison in structural properties between LAEs and other galaxy populations, we use $165517$ photo-$z$ SFGs at $z=0-6$ in \citet{2014ApJS..214...24S} and $10454$ LBGs at $z\simeq4-10$ in \citet{2016ApJ...821..123H}. The $n$ and $r_{\rm e}$ values for these SFGs and LBGs are measured in \citetalias{2015ApJS..219...15S} in the same method as those for LAEs (Section \ref{sec_analysis}). This ensures a fair comparison in the structural quantities. In \citetalias{2015ApJS..219...15S}, the S\'ersic index has not been obtained for the LBGs because the $n$ parameter is fixed for a fair comparison with previous studies on LBGs. In this study, we calculate the S\'ersic index for the LBGs to compare with LAEs at a high redshift of $z\simeq6-7$. 

We also use the $n$ and $r_{\rm e}$ measurements for LAEs in individual previous studies of \citet{2015AA...576A..51G},  \citet{2016ApJ...817...79H}, \citet{2011ApJ...743....9G}, \citet{2012ApJ...753...95B}, \citet{2016ApJ...819...25K}, \citet{2009ApJ...701..915T}, \citet{2012ApJ...750L..36M}, \citet{2013ApJ...773..153J}, \citet{2018MNRAS.476.5479P}. Table \ref{tab_previous_studies} summarizes the redshift range, the number of LAEs, and the method to measure structural quantities for these references including statistical studies on the galaxy morphology \citep[i.e., ][]{2007ApJ...667...49P, 2014ApJ...785...64S, 2014ApJ...786...59H, 2016AA...587A..98W, 2017ApJ...838....4Y, 2017MNRAS.468.1123S, 2017AA...608A...8L}. The effective radius is derived in different techniques or softwares such as {\tt GALFIT}, {\tt GALAPAGOS} \citep{2012MNRAS.422..449B}, {\tt SExtractor} \citep{1996A&AS..117..393B}, or {\tt PHOT}/{\tt ELLIPSE} in {\tt IRAF}, noted in a column of Table \ref{tab_previous_studies}. Among the previous studies, \citet{2015AA...576A..51G}, \citet{2016ApJ...817...79H}, \citet{2012ApJ...753...95B}, \citet{2011ApJ...743....9G}, \citet{2009ApJ...701..915T}, and \citet{2016ApJ...819...25K} provide the structural quantities of $n$ and $r_{\rm e}$ with $M_{\rm UV}$ for individual sources in readable tables. Using the tables, we re-calculate representative values of $n$ and $r_{\rm e}$ in the $M_{\rm UV}$ bins.

\begin{figure}[t!]
  \begin{center}
    \includegraphics[width=85mm]{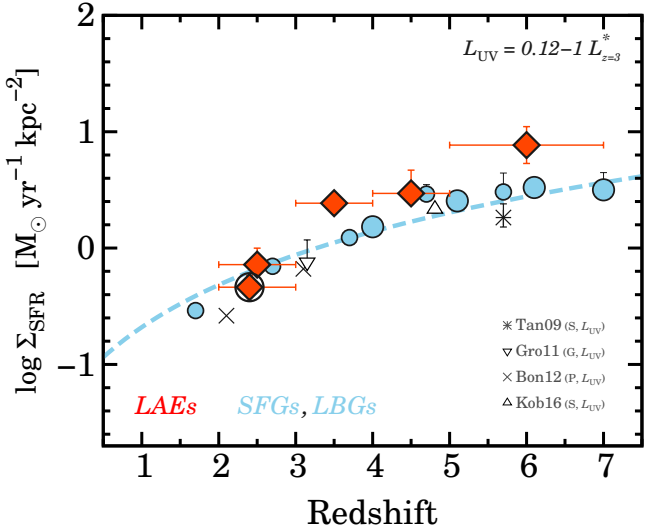}
  \end{center}
  \caption[]{{\footnotesize Same as Figure \ref{fig_z_re_lae}, but for the SFR SD, $\Sigma_{\rm SFR}$. The cyan dashed curve represents the $\Sigma_{\rm SFG}$ evolution calculated with an SFR of $5\,M_\odot/$yr and the best-fit $r_{\rm e}$ function in Figure \ref{fig_z_re_lae}. In contrast to \citetalias{2015ApJS..219...15S}, the SFR for the LBGs is not corrected for dust extinction for a fair comparison with the LAEs. }}
  \label{fig_z_sfrsd_lae}
\end{figure}

\begin{deluxetable}{cccccc}
\setlength{\tabcolsep}{0.35cm} 
\tabletypesize{\scriptsize}
\tablecaption{Size-Luminosity Relation for LAEs}
\tablehead{\colhead{$M_{\rm UV}$} & \colhead{$r_{\rm e}^{\rm Opt}$} & \colhead{$M_{\rm UV}$} & \colhead{$r_{\rm e}^{\rm UV}$} &\colhead{$M_{\rm UV}$} & \colhead{$r_{\rm e}^{\rm UV}$}\\
\colhead{[mag]}& \colhead{[kpc]}& \colhead{[mag]}& \colhead{[kpc]}&\colhead{[mag]}& \colhead{[kpc]}\\
\colhead{(1)}& \colhead{(2)}& \colhead{(3)}& \colhead{(4)} &  \colhead{(5)}& \colhead{(6)} } 

\startdata

\multicolumn{2}{c}{$z=0-1$} & \multicolumn{2}{c}{$z\simeq2$} & \multicolumn{2}{c}{$z\simeq4$} \\
$-19.0$ & $1.44^{+0.148}_{-0.155}$ & $-20.0$ & $0.990^{+0.102}_{-0.042}$ &  $-20.0$ & $0.546^{+0.065}_{-0.041}$ \\
$-17.0$ & $1.02^{+0.508}_{-0.152}$ & $-18.0$ & $0.925^{+0.065}_{-0.030}$ & $-18.0$ & $0.481^{+0.072}_{-0.043}$\\
\multicolumn{2}{c}{$z\simeq2$} & \multicolumn{2}{c}{$z\simeq3$} & \multicolumn{2}{c}{$z\simeq5$} \\
$-20.0$ & $1.11^{+0.092}_{-0.062}$ & $-20.0$ & $0.705^{+0.081}_{-0.065}$ & $-21.0$ & $0.505^{+0.369}_{-0.044}$ \\
$-18.0$ & $0.938^{+0.073}_{-0.040}$ & $-18.0$ & $0.387^{+0.104}_{-0.036}$ & $-19.0$ & $0.466^{+0.042}_{-0.035}$ \\
\nodata & \nodata & \nodata & \nodata  & \multicolumn{2}{c}{$z\simeq6$} \\
\nodata & \nodata & \nodata & \nodata  & $-22.0$ & $0.785^{+0.221}_{-0.221}$ \\
\nodata & \nodata & \nodata & \nodata  & $-20.0$ & $0.543^{+1.64}_{-0.203}$

\enddata

\tablecomments{Columns: (1) (3) (5) UV magnitude. (2) (4) (6) Median effective radius and its standard error at the rest-frame optical or UV wavelengths. }
\label{tab_muv_re}
\end{deluxetable}

\section{RESULTS}\label{sec_result}

We present results of our structure analyses: the S\'ersic index (Section \ref{sec_result_rad}), the $r_{\rm e}$ distribution (Section \ref{sec_result_dist}), the size-luminosity relation (Section \ref{sec_result_muv_re}), the size evolution (Section \ref{sec_result_size_evolution}), and the star formation rate surface density (SFR SD; Section \ref{sec_result_sfrsd}). Similar to \citetalias{2015ApJS..219...15S}, results are shown in a UV luminosity bin of $0.12-1$ $L_{\rm UV}/L_{z=3}^*$. For the representative values and uncertainties of these structural quantities, we employ the medians and standard errors, respectively, unless otherwise specified. For the previous studies on LAEs, we re-calculate the medians and standard errors if structural quantities of individual sources are available (see Section \ref{sec_comparison}).

\subsection{S\'ersic Index}\label{sec_result_rad}

The left panel of Figure \ref{fig_rad_prof_all} shows the S\'ersic index as a function of redshift. We find that LAEs have a nearly constant $n$ value of $n\simeq1-1.5$ in a wide redshift range of $z\simeq2-7$. There is no significant difference in $n$ for LAEs at $z\simeq2$ between the rest-frame UV and optical wavelengths. This constant and low $n$ value is comparable to the $n$ trend for the SFGs and LBGs. The $n$ result in our LAE sample are consistent with that in previous studies on LAEs, \citet{2009ApJ...701..915T} and \citet{2011ApJ...743....9G}. \citet{2018MNRAS.476.5479P}'s S\'ersic index is slightly higher than our and these previous studies' results by $\Delta n\simeq0.5-1$, but is still lower than $n\simeq2-3$. We confirm that LAEs show no significant evolution in S\'ersic index and a low $n$ value of $n\simeq1-1.5$, previously reported in \citet{2018MNRAS.476.5479P}. This low S\'ersic index indicates that LAEs typically have a disk-like SB profile in the stellar continuum emission. 

Complementary to the $n$ measurements for the individual sources, we also investigate the radial SB profiles in the stacked galaxy images. In Figure \ref{fig_rad_prof_all}, we demonstrate that the radial SB profiles are similar to the exponential function (i.e., $n=1$) instead of de Vaucouleurs' law (i.e., $n=4$). The stacked galaxy images are created in a method similar to that in \citetalias{2016ApJ...821...72S}. We fit the 1D S\'ersic functions with $n=1$ and $n=4$ to the radial SB profiles. The data points at $r>0.2$ arcsec are used for the fitting to avoid the PSF broadening effect (\citetalias{2016ApJ...821...72S}). As shown at the galactic central regions of $r<0.2$ arcsec, the 1D S\'ersic function with $n=4$ deviates from the radial SB profiles of the LAEs, SFGs, and LBGs. Based on the fitting results at the high-$S/N$ central regions, we conclude that our SB profiles are better represented by the function of $n=1$ than $n=4$. The analysis of the stacked {\it HST} images supports the measurements of $n\simeq1$ for individual LAEs. These radial SB profiles in the stacked {\it HST} images are analyzed for a comparison with Ly$\alpha$ halos in Section \ref{sec_discuss_lya_halo}.

\subsection{Distribution of the effective radius}\label{sec_result_dist}

We investigate the distribution of the effective radius, $r_{\rm e}$, and its width. According to the galaxy disk formation models of e.g., \citet{1980MNRAS.193..189F}, \citet{1987ApJ...319..575B}, \citet{1992ApJ...399..405W}, \citet{1998MNRAS.295..319M}, \citet{2001ApJ...555..240B} and results of DM N-body simulations, the DM spin parameter $\lambda$ is predicted to have a log-normal distribution with the standard deviation of $\sigma_{\ln{\lambda}}\sim0.5-0.6$. If the stellar components of LAEs form in such disk formation mechanisms and $r_{\rm e}$ is determined by $\lambda$, the $r_{\rm e}$ distribution should be log-normal and $\sigma_{\ln{r_{\rm e}}}\simeq0.5-0.6$ similar to $\lambda$. Recently, the disk formation models have been tested with observational $r_{\rm e}$ data of galaxies at $z\simeq0-3$ \citep[e.g., ][]{2017arXiv171110500H, 2017ApJ...838....6H, 2018ApJ...854...22O, 2018MNRAS.473.2714S, 2018arXiv180802525F}. 

Figure \ref{fig_hist_re_L012_1_lae_all} presents the $r_{\rm e}^{\rm UV}$ distribution of our LAEs in $L_{\rm UV}=0.12-1\,L_{z=3}^*$. The LAEs appear to be represented by a log-normal distribution albeit with poor statistics especially for $z\simeq4-7$. The $r_{\rm e}$ distribution is fitted by the log-normal function of 

\begin{equation}\label{eq_lognormal}
 p(r_{\rm e}) = \frac{1}{r_{\rm e}\sigma_{\ln{r_{\rm e}}}\sqrt{2\pi}} \exp \biggl[ -\frac{\ln^2 (r_{\rm e}/\overline{r_{\rm e}})}{2\sigma^2_{\ln{r_{\rm e}}}} \biggr]
\end{equation}

\noindent where $\overline{r_{\rm e}}$ and $\sigma_{\ln{r_{\rm e}}}$ are the peak of $r_{\rm e}$ and the standard deviation of $\ln{r_{\rm e}}$, respectively. We obtain the best-fit distribution width of $\sigma_{\ln{r_{\rm e}}} = 0.72\pm0.06$ at $z\simeq2-3$, $0.59\pm0.02$ at $z\simeq3-4$, and $0.42\pm0.10$ at $z\simeq4-7$ corrected for the uncertainty of $r_{\rm e}$ measurements. These $\sigma_{\ln{r_{\rm e}}}$ values are comparable to $\sigma_{\ln{r_{\rm e}}}\simeq0.45-0.75$ of the SFGs and LBGs estimated in \citetalias{2015ApJS..219...15S} and the width of the DM spin parameter, $\sigma_{\ln{r_{\rm e}}}\simeq0.5-0.6$. This result might imply that the stellar components of LAEs form in the galaxy disk formation models. The relation between the disk formation and the LAE morphology is discussed in Section \ref{sec_discuss_disk} in details.

\subsection{Size-Luminosity Relation}\label{sec_result_muv_re}

We explore the size-luminosity, $r_{\rm e}$-$L_{\rm UV}$, relation. Figure \ref{fig_muv_re_lae_all} shows the $r_{\rm e}$-$L_{\rm UV}$ relation for the LAEs, SFGs, and LBGs, where $L_{\rm UV}$ is presented with the corresponding $M_{\rm UV}$. We find a trend that $r_{\rm e}$ is larger at a brighter $L_{\rm UV}$ (i.e., smaller $M_{\rm UV}$) for the LAEs, similar to the SFGs and LBGs. Although the some median $r_{\rm e}$ slightly deviate from the $r_{\rm e}$-$L_{\rm UV}$ relation of the SFGs and LBGs (e.g., at $z\simeq5$), the data points for the LAEs are consistent with those of the SFGs and LBGs within a $1\sigma$ error. This trend is shown in both of the rest-frame UV and optical wavelengths (i.e., the top and bottom panels in Figure \ref{fig_muv_re_lae_all}). Figure \ref{fig_muv_re_lae_all} also provides the best-fit $r_{\rm e}$-$L_{\rm UV}$ relation obtained in \citetalias{2015ApJS..219...15S} for the SFGs and LBGs:  

\begin{equation}\label{eq_re_luv}
r_{\rm e} = r_0 \Biggl( \frac{L_{\rm UV}}{L_0} \Biggr) ^\alpha, 
\end{equation} 

\noindent where $r_0$ and $\alpha$ are free parameters. The $L_0$ values are employed as the typical UV luminosity at $M_{\rm UV}=-21$. Our LAEs approximately follow the best-fit $r_{\rm e}$-$L_{\rm UV}$ relation of the SFGs and LBGs. Each $r_{\rm e}$ value of the LAEs is listed in Table \ref{tab_muv_re}. 

We compare our measurements with that of previous studies on LAEs, \citet{2009ApJ...701..915T}, \citet{2011ApJ...743....9G}, \citet{2012ApJ...753...95B}, \citet{2013ApJ...773..153J}, \citet{2015AA...576A..51G}, \citet{2016ApJ...817...79H}, and \citet{2016ApJ...819...25K}. Similar to our LAE sample, LAEs in these previous studies are located around the $r_{\rm e}$-$L_{\rm UV}$ relation of the SFGs and LBGs.

\subsection{Size Evolution}\label{sec_result_size_evolution}

We examine the redshift evolution in galaxy sizes. In this study, we discuss the evolutional trend of median $r_{\rm e}$ which is a proxy for the typical galaxy size of our samples. Figure \ref{fig_z_re_lae} plots the effective radius as a function of redshift for our LAE, SFG, and LBG samples. As shown in Figure \ref{fig_z_re_lae}, the LAEs significantly evolve in $r_{\rm e}$ at the rest-frame UV wavelength from $r_{\rm e}^{\rm UV}\simeq0.5$ kpc at $z\simeq4-6$ to $r_{\rm e}^{\rm UV}\simeq1$ kpc at $z\simeq2$, similarly found for the SFGs and LBGs. We parametrize the size growth rate by fitting a function of $r_{\rm e}(z) = B (1+z)^\beta$, where $B$ and $\beta$ are free parameters. The fitting is performed using only the three $r_{\rm e}^{\rm UV}$ values from $z\simeq2$ to $z\simeq7$. We find that $r_{\rm e}^{\rm UV}$ scales as $\propto (1+z)^{-1.37\pm0.65}$, indicating that there is a size evolution similar to the SFGs and LBGs with $\beta\simeq-1\sim-1.5$ (\citetalias{2015ApJS..219...15S}). This result is consistent with the evolution of the size-luminosity relation for the LAEs, SFGs, and LBGs, as shown in Section \ref{sec_result_muv_re}. The galaxy size at the rest-frame optical wavelength, $r_{\rm e}^{\rm Opt}$, follows the evolution of $r_{\rm e}^{\rm UV}$ for the LAEs at $z\simeq2$, suggesting, again, that the choice of the observed wavelength gives no significant impact on the galaxy size. In addition, we plot $r_{\rm e}^{\rm Opt}$ at $z\simeq0-1$ in Figure \ref{fig_z_re_lae}. Although there are a few LAEs at $z\simeq0-1$ falling in the $L_{\rm UV} = 0.12-1 L_{z=3}^*$ bin, we infer $r_{\rm e}^{\rm Opt}$ by extrapolating the size-luminosity relation in Figure \ref{fig_muv_re_lae_all}. The $r_{\rm e}^{\rm Opt}$ is $\simeq2$ kpc at $z\simeq0-1$, supporting the size evolution of LAEs from $r_{\rm e}\simeq0.5$ kpc at $z\simeq4-7$. 

Figure \ref{fig_z_re_lae} compares the effective radii for our and previous studies on LAEs, \citet{2009ApJ...701..915T}, \citet{2011ApJ...743....9G}, \citet{2012ApJ...753...95B}, \citet{2012ApJ...750L..36M}, \citet{2015AA...576A..51G}, \citet{2016ApJ...817...79H}, \citet{2016ApJ...819...25K}, and \citet{2018MNRAS.476.5479P}. Note that the median and the error bars of $r_{\rm e}$ in $L_{\rm UV} = 0.12-1 L_{z=3}^*$ are re-calculated for previous studies which give the information of $r_{\rm e}$ and $M_{\rm UV}$ for individual sources. As found in Figure \ref{fig_z_re_lae}, most $r_{\rm e}$ measurements of previous studies are broadly consistent with the best-fit $r_{\rm e}(z)$ function obtained in our LAE sample. The effective radius of \citet{2009ApJ...701..915T}, \citet{2012ApJ...750L..36M}, and \citet{2018MNRAS.476.5479P} is slightly higher than our $r_{\rm e}$ measurements. Among the studies with $r_{\rm e}$ slightly higher than ours, \citet{2012ApJ...750L..36M} and \citet{2018MNRAS.476.5479P} do not provide the $r_{\rm e}$ and $M_{\rm UV}$ table, which is unable us to re-calculate the median $r_{\rm e}$ in the $L_{\rm UV}$ range. In Section \ref{sec_discuss}, we discuss the difference in the galaxy size growth rates for LAEs between previous studies and ours.

\begin{figure*}[t!]
  \begin{center}
    \includegraphics[width=85mm]{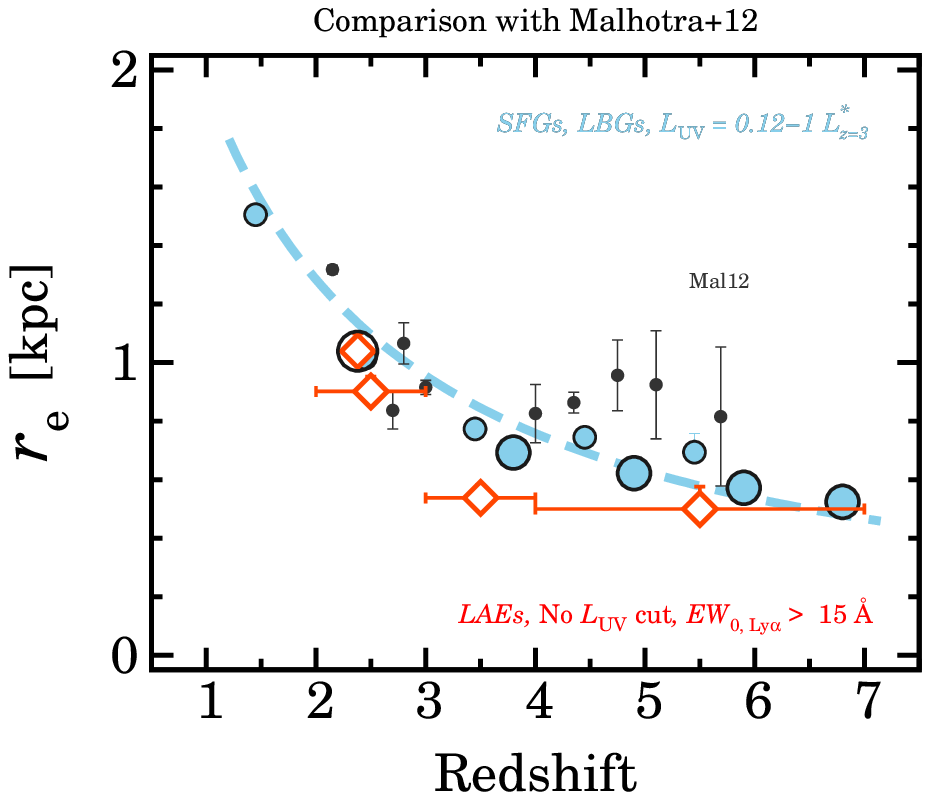} \mbox{}
    \includegraphics[width=85mm]{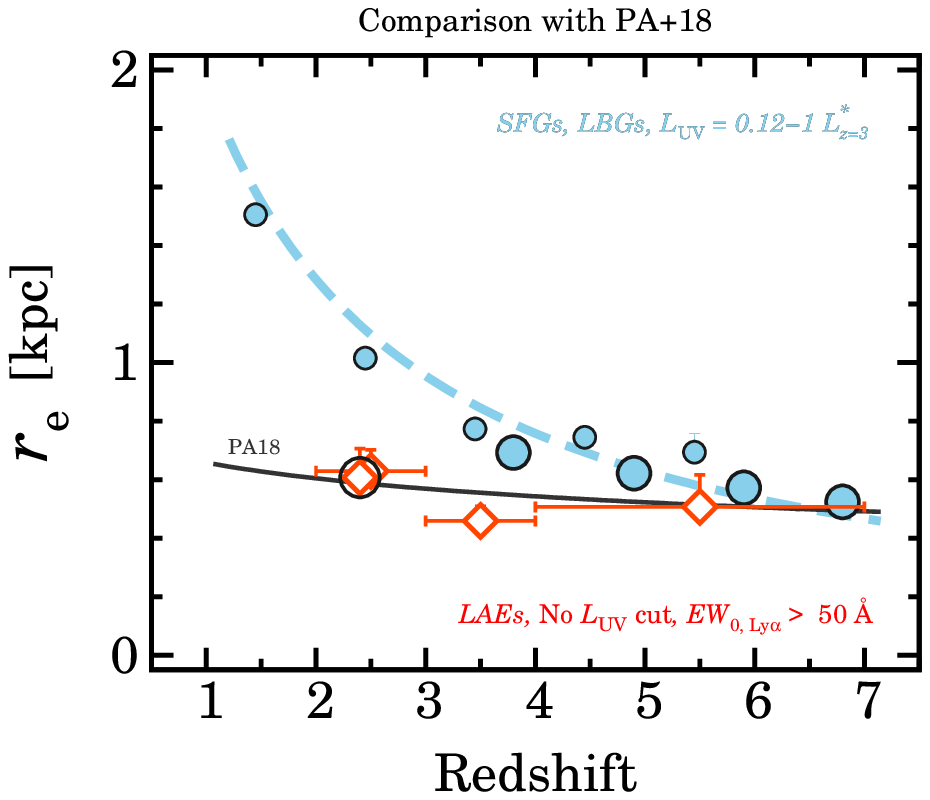}
  \end{center}
  \caption[]{{\footnotesize Same as Figure \ref{fig_z_re_lae}, but for a comparison with the size evolution with \citet{2012ApJ...750L..36M} (left panel) and \citet{2018MNRAS.476.5479P} (right panel). The black filled circles indicate the $r_{\rm e}$ measurements in \citet{2012ApJ...750L..36M}. The black solid line denotes the no size evolution curve of $r_{\rm e}\propto (1+z)^{-0.21}$ in \citet{2018MNRAS.476.5479P} which is matched to our $r_{\rm e}$ data point for LAEs at $z=4-7$. The open diamonds represent our LAEs that are re-sampled with $EW_{\rm 0, Ly\alpha}>15$\,\AA\, (left panel) and $EW_{\rm 0, Ly\alpha}>50$\,\AA (right panel). In both cases, no $L_{\rm UV}$ cut is adopted. }}
  \label{fig_z_re_lae_ewlya}
\end{figure*}

\subsection{SFR Surface Density}\label{sec_result_sfrsd}

Figure \ref{fig_z_sfrsd_lae} presents the redshift evolution of SFR SD, $\Sigma_{\rm SFR}$, which is one of the size-relevant quantities. The SFR SD is derived with $r_{\rm e}$ by 

\begin{equation}\label{eq_sfrsd}
\Sigma_{\rm SFR} \, {\rm [M_\odot\, yr^{-1}\,{\rm kpc}^{-2}]} = \frac{{\it SFR_{\rm UV}}/2}{\pi r_{\rm e}^2}. 
\end{equation}

We compute SFRs at the rest-frame UV wavelength from $L_{\rm UV}$ using the relation of \citet{1998ARA&A..36..189K},

\begin{equation}\label{eq_sfr}
 {\it SFR}_{\rm UV} \,{\rm [M_\odot \, yr^{-1}]} = 1.4 \times 10^{-28} L_\nu \, [{\rm erg\, s}^{-1}\,{\rm Hz}^{-1}]. 
\end{equation}

\noindent For all the populations, we do not take into account the dust extinction, $E(B-V)$, due to the difficulty in estimating $E(B-V)$ from e.g., spectral slopes of the rest-frame UV continuum emission, for typically faint LAEs. 

As shown in Figure \ref{fig_z_sfrsd_lae}, $\Sigma_{\rm SFR}$ for the SFGs and LBGs gradually increases from $z\simeq 1$ to $z\simeq7$. In Figure \ref{fig_z_sfrsd_lae}, we depict the $\Sigma_{\rm SFR}$ evolution curve of the SFGs and LBGs using Equation (\ref{eq_sfrsd}) with the inputs of the best-fit function of $r_{\rm e}(z)$ in Section \ref{sec_result_sfrsd} and the SFR estimated from the $L_{\rm UV}$ value via Equation (\ref{eq_sfr}). We find that the LAEs in our and previous studies \citep{2009ApJ...701..915T, 2011ApJ...743....9G, 2012ApJ...753...95B, 2016ApJ...819...25K} follows the $\Sigma_{\rm SFR}$ evolutional trend.

\section{DISCUSSION} \label{sec_discuss}

\subsection{LAE size evolves or not?} \label{sec_discuss_size_evolution}

One of the most striking features that are different from results in previous studies is the size growth rate for LAEs. Our analysis has suggested that LAEs evolve in $r_{\rm e}$ similar to SFGs and LBGs (Section \ref{sec_result_size_evolution} and Figure \ref{fig_z_re_lae}). In contrast, some previous studies have reported that the $r_{\rm e}$ of LAEs is almost constant over the cosmic time of $z\simeq2-6$ (e.g., \citealt{2012ApJ...750L..36M}, \citealt{2018MNRAS.476.5479P}). A potential source making the difference is the binning of the galaxy luminosity (or galaxy mass) in galaxy samples. For a fair $r_{\rm e}$ comparison, an identical $L_{\rm UV}$ range should be applied to 1) the comparison samples (SFGs and LBGs in this case) and 2) LAEs. As shown below, the $r_{\rm e}$ comparison in the previous studies is probably affected by the $L_{\rm UV}$ heterogeneity. 

To demonstrate the importance of the $L_{\rm UV}$ binning, we compare the size growth rate of our LAE sample with that of \citet{2012ApJ...750L..36M} and \citet{2018MNRAS.476.5479P} by matching the $EW_{\rm 0, Ly\alpha}$ and $L_{\rm UV}$ ranges. The left panel of Figure \ref{fig_z_re_lae_ewlya} shows the result of \citeauthor{2012ApJ...750L..36M} We re-sample our LAEs with a Ly$\alpha$ EW limit of $EW_{\rm 0, Ly\alpha}>15$\AA\, and without a $L_{\rm UV}$ cut which are similar to the selection criteria of \citeauthor{2012ApJ...750L..36M} In these $EW_{\rm 0, Ly\alpha}$ and $L_{\rm UV}$ ranges, we find that the $r_{\rm e}$ evolution of our re-sampled LAEs broadly agrees with that of \citeauthor{2012ApJ...750L..36M} Interestingly, \citeauthor{2012ApJ...750L..36M}'s data points appear to follow the size evolution curve of the SFGs and LBGs, contrary to \citeauthor{2012ApJ...750L..36M}'s report. According to the $r_{\rm e}$-$L_{\rm UV}$ relation of SFGs and LBGs \citep[e.g., \citetalias{2015ApJS..219...15S}; ][]{2012ApJ...756L..12M, 2013ApJ...777..155O, 2014ApJ...788...28V}, the size evolution curve is shifted along the $r_{\rm e}$ direction with changing a $L_{\rm UV}$ range. In the case that the $L_{\rm UV}$ range is not matched, one could identify a false difference in the size evolution curves between the comparison samples and LAEs. Thus, the conclusion of the no $r_{\rm e}$ evolution in \citeauthor{2012ApJ...750L..36M} would be caused mostly by the $L_{\rm UV}$ difference between LAEs and comparison samples. 

The right panel of Figure \ref{fig_z_re_lae_ewlya} shows the result of \citeauthor{2018MNRAS.476.5479P} We re-sample our LAEs with a Ly$\alpha$ EW limit of $EW_{\rm 0, Ly\alpha}>50$\AA\, and without a $L_{\rm UV}$ cut which are similar to the selection criteria of \citeauthor{2018MNRAS.476.5479P} As shown in the right panel of Figure \ref{fig_z_re_lae_ewlya}, we reproduce a nearly constant effective radius at $r_{\rm e}\simeq0.5$ kpc that matches well the size evolution curve of \citeauthor{2018MNRAS.476.5479P} However, we find a selection bias in this re-sampled LAEs and \citeauthor{2018MNRAS.476.5479P}'s sample (see Figure \ref{fig_z_luv} and \citeauthor{2018MNRAS.476.5479P}'s Table 1). In general astronomical surveys, faint sources tend to be more detectable than bright ones at low-$z$. According to the $r_{\rm e}$-$L_{\rm UV}$ relation, one could see galaxy sizes that are biased to small $r_{\rm e}$ at low-$z$ and large $r_{\rm e}$ at high-$z$, if no $L_{\rm UV}$ binning is applied. Moreover, the relatively high Ly$\alpha$ EW limit of $EW_{\rm 0, Ly\alpha}>50$\AA\, tends to exclude $L_{\rm UV}$-brighter LAEs at lower-$z$ due to the $EW_{\rm 0, Ly\alpha}$-$L_{\rm UV}$ anti-correlation \citep{2006ApJ...645L...9A}, more severely enhancing the selection bias. Such a different $L_{\rm UV}$ range would compensate the intrinsic size evolution of LAEs. 

For these reasons, the size evolution should be compared at a given luminosity. In our study, $L_{\rm UV}$ is confined to $L_{\rm UV}=0.12-1 L_{z=3}^*$ for a fair $r_{\rm e}$ comparison between our samples of LAEs, SFGs, and LBGs. In this $L_{\rm UV}$ range, we find that LAEs show the size evolution even in the literature LAE samples (i.e., data points of \citealt{2011ApJ...743....9G}, \citealt{2012ApJ...753...95B} and \citealt{2016ApJ...817...79H} in Figure \ref{fig_z_re_lae}). In addition to the $L_{\rm UV}$ binning, note that we match the choice of statistics (i.e. the median), the technique to measure $r_{\rm e}$ (i.e. {\tt GALFIT}), and the band of the {\it HST} images to trace the stellar continuum emission (see Section \ref{sec_analysis}). Our systematic analysis has identified the size evolution for LAEs similar to SFGs and LBGs.

\begin{figure}[t!]
  \begin{center}
    \includegraphics[width=85mm]{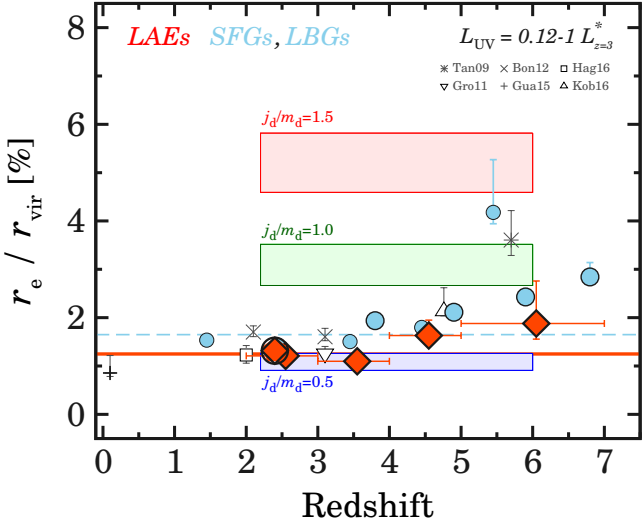}
  \end{center}
  \caption[]{{\footnotesize Same as Figure \ref{fig_z_re_lae}, but for the SHSR, $r_{\rm e}/r_{\rm vir}$. The magenta solid and cyan dashed horizontal lines indicate weighted means of $\left< r_{\rm e}/r_{\rm vir} \right>$ for the LAEs and for the SFGs and LBGs, respectively. The virial mass of host DM halos is derived from the results of \citet{2013ApJ...770...57B}. The red, green, and blue shaded areas illustrate the regions of $j_{\rm d}/m_{\rm d}=1.5$, $1.0$, and $0.5$, respectively (see Section \ref{sec_discuss_disk} for details). }}
  \label{fig_z_rervir_lae}
\end{figure}

\subsection{Implications for the galaxy disk formation of LAEs} \label{sec_discuss_disk}

As indicated by the low S\'ersic index of $n\simeq1-1.5$ in Section \ref{sec_result_rad}, we have revealed that LAEs typically have a disk-like SB profile in the stellar continuum emission. In addition, the $r_{\rm e}$ distribution and its width are consistent with a picture that stellar components of LAEs is formed in host DM halos through galaxy disk formation models (Section \ref{sec_result_dist}). To provide the implications of the galaxy disk formation of LAEs, we infer the stellar-to-halo size ratios (SHSRs) that are defined as the ratio of $r_{\rm e} / r_{\rm vir}$, where $r_{\rm vir}$ the virial radius of a host DM halo. 

Figure \ref{fig_z_rervir_lae} plots $r_{\rm e}/r_{\rm vir}$ as a function of redshift in the luminosity range of $L_{\rm UV}=0.12-1 L^*_{\rm z=3}$. The technique to estimate the SHSRs is the same as that in \citet{2015ApJ...804..103K} and \citetalias{2015ApJS..219...15S}. The $r_{\rm vir}$ value is calculated by 

\begin{equation}\label{eq_virial}
  r_{\rm vir} = \Biggl(  \frac{2GM_{\rm vir}}{\Delta_{\rm vir}\Omega_{\rm m}(z)H(z)^2} \Biggr)^{1/3}, 
\end{equation}

\noindent where $\Delta_{\rm vir}=18\pi^2+82x-39x^2$ and $x=\Omega_m (z) - 1$ \citep{1998ApJ...495...80B}. We obtain the virial mass of a DM halo, $M_{\rm vir}$, from stellar mass, $M_*$, of individual galaxies by using the relation determined by the abundance matching analyses \citep{2010ApJ...717..379B,2013ApJ...770...57B}. In the $M_*$ range for our LAEs (i.e., $\log{M_*/M_\sun}\simeq9-10$), the assumed $M_*/M_{\rm vir}$ ratio varies by $\simeq\pm0.3-0.4$ dex, depending on results of theoretical and observational studies \citep[e.g., ][]{2018arXiv180607893B}. Given the weak dependence of $M_{\rm vir}$ on $r_{\rm vir}$ (Equation \ref{eq_virial}), the assumed $M_*/M_{\rm vir}$ ratio does not significantly affect $r_{\rm e}/r_{\rm vir}$ and our conclusion. To estimate the stellar mass $M_*$, we use $M_{\rm UV}$ and the empirical $M_*$-$M_{\rm UV}$ relation in \citetalias{2015ApJS..219...15S}.  

In Figure \ref{fig_z_rervir_lae}, we find that the $r_{\rm e}/ r_{\rm vir}$ estimates for LAEs fall within the range of $r_{\rm e}/ r_{\rm vir}=1-2$\% similar to the SFGs and LBGs. Interestingly, $r_{\rm e}/r_{\rm vir}$ of the LAEs is nearly constant with redshift, albeit with the large uncertainties at $z\gtrsim5$. The weighted means of the $r_{\rm e}/ r_{\rm vir}$ measurements at $\lambda_{\rm UV}$ are $1.25\pm0.09$\% for the LAEs and $1.65\pm0.10$\% for the SFGs and LBGs. For comparison, we apply this calculation to the literature LAE samples with $L_{\rm UV}$ \citep[i.e., ][]{2009ApJ...701..915T, 2011ApJ...743....9G, 2012ApJ...753...95B, 2015AA...576A..51G, 2016ApJ...817...79H, 2016ApJ...819...25K}. As shown in Figure \ref{fig_z_rervir_lae}, our LAEs agree in $r_{\rm e}/ r_{\rm vir}$ with these literature LAE samples. 

According to the galaxy disk formation models of e.g., \citet{1980MNRAS.193..189F} and \citet{1998MNRAS.295..319M}, the SHSR is liked to the specific angular momentum, $j_{\rm d}/m_{\rm d}$. Figure \ref{fig_z_rervir_lae} also shows $r_{\rm e}/r_{\rm vir}$ regions corresponding to $j_{\rm d}/m_{\rm d}=0.5$, $1.0$, and $1.5$ inferred from 

\begin{equation}\label{eq_spin}
  \frac{r_{\rm e}}{r_{\rm vir}} = \frac{1.678}{\sqrt{2}} \Biggl(\frac{j_{\rm d}}{m_{\rm d}}\lambda \Biggr) \frac{f_{\rm R}(\lambda, c_{\rm vir}, m_{\rm d}, j_{\rm d})}{\sqrt{f_c(c_{\rm vir})}}, 
\end{equation}

\noindent where the $j_{\rm d}$ ($m_{\rm d}$) value is a angular momentum (mass) ratio of a central disk to a host DM halo. The $f_c(c_{\rm vir})$ and $f_{\rm R}(\lambda, c_{\rm vir}, m_{\rm d}, j_{\rm d})$ are functions related to halo and baryon concentrations, respectively. The $c_{\rm vir}$ is the halo concentration factor. If we use $\lambda$ and $c_{\rm vir}$ values well constrained by numerical simulations \citep[e.g., ][]{2002ApJ...581..799V,2009MNRAS.393.1498D,2012MNRAS.423.3018P}, we can constrain $j_{\rm d}/m_{\rm d}$. Figure \ref{fig_z_rervir_lae} presents that our estimates of $r_{\rm e}/r_{\rm vir}$ range from $j_{\rm d}/m_{\rm d}\simeq0.5$ to $\simeq1$. This result of $j_{\rm d}/m_{\rm d}\sim0.5-1$ indicates that a central galaxy of LAEs acquire more than half of specific angular momentum from a host DM halo. 

As presented in the previous sections, our systematic structural analyses have revealed that LAEs have the S\'ersic index, the $r_{\rm e}$ distribution, the size growth rate, the SFR SD, and the SHSR that are comparable with those of SFGs and LBGs. These morphological similarities between LAEs and other galaxy populations have already been reported at $z\simeq2-3$ \citep[e.g., ][]{2016ApJ...817...79H, 2017MNRAS.468.1123S}. We confirm this similarity in the wide redshift range of $z\simeq0-7$ in our large LAE sample. The results naturally indicate that the mechanism of the Ly$\alpha$ photon escape is related to e.g., properties of the interstellar/circum-galactic medium, e.g., the amount \citep[e.g., ][]{2014ApJ...794..101P, 2014ApJ...788...74S, 2015ApJ...812..157H}, geometry \cite[e.g., ][]{2013arXiv1308.1405Z},  kinematics \citep[e.g., ][]{2016A&A...587A..78H}, and ionization states \citep[e.g., ][]{2013arXiv1309.0207N}, rather than the global galaxy morphology in the stellar components. In addition to phenomena related to non-stellar physics, the star formation history would be an important factor controlling the Ly$\alpha$ emissivity \citep[e.g., ][]{2003A&A...397..527S}. Spatially-resolved and multi-wavelength analyses need to be performed to understand the relation between stellar populations, galaxy morphological properties, and the Ly$\alpha$ emissivity.

\subsection{Origins of Ly$\alpha$ halos}\label{sec_discuss_lya_halo}

\begin{figure}[t!]
  \begin{center}
    \includegraphics[width=80mm]{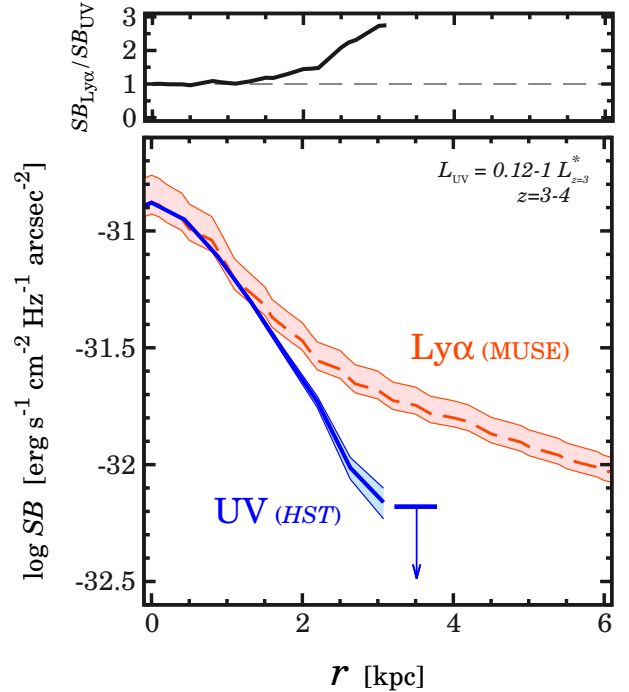}
  \end{center}
  \caption[]{{\footnotesize Radial SB profiles for LAEs at $z=3-4$ in the $L_{\rm UV}$ bin of $L_{\rm UV}=0.12-1 L_{z=3}^*$. The blue solid curve and red dashed curves show the median radial SB profiles derived from the rest-frame UV continuum (this study) and Ly$\alpha$ \citep{2017AA...608A...8L} emission, respectively. The shaded regions depict the standard error of each radial SB profile. The radial SB profile in the rest-frame UV continuum emission is displayed above a $\sim5\sigma$ detection in our {\it HST} stacked image. The radial SB profiles in Ly$\alpha$ and UV are compared in the same PSF size. The top-panel presents the ratio of ${\it SB}_{\rm Ly\alpha}$ and ${\it SB}_{\rm UV}$ as a function of radius. }}
  \label{fig_rad_prof_lae}
\end{figure}

In this last section, we discuss the physical origins of Ly$\alpha$ halos with the {\it HST} data of the stellar components. Using our large LAE sample, we have created radial SB profiles in the rest-frame UV stellar continuum emission in the stacked {\it HST} images (Section \ref{sec_result_rad}). Here we compare our radial SB profile in UV with one in Ly$\alpha$ obtained in a MUSE IFU observation \citep{2017AA...608A...8L}. Figure \ref{fig_rad_prof_lae} presents the radial profiles in UV and the Ly$\alpha$ halo for LAEs at $z=3-4$ in $L_{\rm UV}=0.12-1 L_{z=3}^*$. The Ly$\alpha$ halo is a median stacked 1D profile of LAEs with the best-fit stellar continuum and halo slopes, $rs_{\rm cont}$ and $rs_{\rm halo}$ (see \citealt{2017AA...608A...8L}) which fall within the ranges of $z=3-4$ and $L_{\rm UV}=0.12-1 L_{z=3}^*$. Considered the ambiguity of the Ly$\alpha$ escape fraction, ${\it SB}_{\rm Ly\alpha}$ is scaled to match ${\it SB}_{\rm UV}$ at $r=0$ kpc. 

As shown in Figure \ref{fig_rad_prof_lae}, we find that \textcolor{red}{${\it SB}_{\rm UV}$} is lower than \textcolor{red}{${\it SB}_{\rm Ly\alpha}$} by a factor of at least $\simeq2-3$ at $r\simeq3$ kpc. This is consistent with the reports in the literature \citep{2011ApJ...736..160S, 2012MNRAS.425..878M, 2014MNRAS.442..110M}. This result indicates that the Ly$\alpha$ halo is not originated only from the recombination of the small satellite galaxies and diffuse stellar disks. The high spatial resolution of the {\it HST} images enables us to investigate the central regions of the Ly$\alpha$ halo, less affected by e.g., the sky oversubtraction and flat-fielding effects \citep{2013ApJ...776...75F, 2016MNRAS.457.2318M}. Interestingly, the SB deficit in UV compared to Ly$\alpha$ is shown in the vicinity of the galaxy center, at $r\simeq1-2$ kpc. The high spatial resolution radial profiles would provide important constraints on the radiation budget of the several Ly$\alpha$ halo formation mechanisms \cite[e.g., ][]{2012MNRAS.424.1672D, 2015ApJ...806...46L, 2016ApJ...822...84M,2017ApJ...841...19M,2017ApJ...846...11M}.

\section{SUMMARY and CONCLUSIONS}\label{sec_conclusion}

We investigate the redshift evolution of radial SB profiles and effective radii, $r_{\rm e}$, of the rest-frame UV and optical stellar continua for $9119$ LAEs at $z\simeq0-8$ that are compiled from literature wide-field NB, MB, and MUSE/IFU data. This LAE sample is the largest for studies of LAE galaxy structures in the wide-redshift range of $z\simeq0-8$. Using the deep extra-galactic legacy data of $\!${\it HST}, we measure the structural quantities for the LAEs to make a comparison with $\simeq180,000$ sources of photo-$z$ SFGs and LBGs at $z\simeq0-10$ in \citetalias{2015ApJS..219...15S}. The structural quantities of the LAEs, SFGs, and LBGs are obtained in the same techniques of the size measurement, the same choice of statistics, the same bands of the {\it HST} images, and the same UV-continuum luminosity range, allowing to evaluate the redshift evolution of galaxy structures with no significant systematics. 

The main results of this study are summarized as follows. 

\begin{enumerate}
  \item The analyses of the radial SB profiles in individual and stacked {\it HST} images reveal that the S\'ersic index is almost constant at $n\simeq1-1.5$ for LAEs over the cosmic time. The S\'ersic index measurements suggest that LAEs typically have a disk-like SB profile in the stellar continuum emission independent of redshift. 
  
  \item The $r_{\rm e}$ distribution shape of LAEs is represented by log-normal functions. The standard deviation of the log-normal $r_{\rm e}$ distribution, $\sigma_{\ln{r_{\rm e}}}$, is $\sigma_{\ln{r_{\rm e}}}\simeq0.45-0.75$ for LAEs which is  comparable to that of the DM spin parameter, $\sigma_{\ln{\lambda}}\sim 0.5-0.6$. The similarity of the distribution shapes and standard deviations of $r_{\rm e}$ and $\lambda$ would imply that galaxy sizes of stellar components are associated with the host DM halo kinematics. 
  
  \item The size-luminosity relation of the LAEs monotonically decreases towards high-$z$, following the size-luminosity relations of SFGs and LBGs. The median $r_{\rm e}$ values of the LAEs significantly evolve as $r_{\rm e}\propto(1+z)^{-1.37\pm0.65}$, similar to those of the SFGs and LBGs in the same luminosity range of $L_{\rm UV} = 0.12-1 L^*_{z=3}$. This size growth rate is in contrast with the claims of no evolution made by previous studies whose LAE samples are probably biased to faint sources at low redshift. 
  
  \item Combining our stellar $r_{\rm e}$ measurements with host DM halo radii, $r_{\rm vir}$, estimated from a abundance matching study, we obtain a median value of $r_{\rm e}/r_{\rm vir}=1-3$\%, which is nearly constant at $z\simeq2-7$, similar to SFGs and LBGs. If we assume the disk formation model of \citet{1998MNRAS.295..319M}, our $r_{\rm e}/r_{\rm vir}$ estimates indicate that a central LAE acquires more than a half of specific angular momentum from their host DM halo,  $j_{\rm d}/m_{\rm d}\simeq 0.5-1$.
  
    \item We compare stacked radial SB profiles in the rest-frame UV and Ly$\alpha$ \citep{2017AA...608A...8L}, ${\it SB}_{\rm UV}$ and ${\it SB}_{\rm Ly\alpha}$. The comparison reveals that ${\it SB}_{\rm UV}$ is lower than ${\it SB}_{\rm Ly\alpha}$ by a factor of $\gtrsim2-3$ at $r\simeq3$ kpc from the galaxy center. This deficit in ${\it SB}_{\rm UV}$ suggests that the flux contribution of satellite small galaxies is $\lesssim30-50$\% to the Ly$\alpha$ halos, which is consistent with estimates of ground-based telescopes. Interestingly, this deficit is found in the vicinity of the galaxy center, at $r\simeq1-2$ kpc. The high spatial resolution radial profiles of the stellar continuum would provide important constraints on the Ly$\alpha$ halo formation mechanisms. 
    
 \end{enumerate}
 
Our structural analyses have revealed that the S\'ersic index, the $r_{\rm e}$ distribution, the size growth rate, the SFR SD, and the SHSR of the LAEs are comparable with those of the SFGs and LBGs. These morphological similarities between stellar components of the LAEs, SFGs, and LBGs would indicate that the Ly$\alpha$ photon escape is controlled by the non-stellar physics such as geometry, kinematics, and ionization states of the inter-stellar/circum-galactic medium. 

\acknowledgments

We thank the anonymous referee for constructive comments and suggestions. This work is based on observations taken by the 3D-HST Treasury Program (GO 12177 and 12328) and CANDELS Multi-Cycle Treasury Program with the NASA/ESA HST, which is operated by the Association of Universities for Research in Astronomy, Inc., under NASA contract NAS5-26555. Support for this work was provided by NASA through an award issued by JPL/Caltech. This work is supported by KAKENHI 16J07046 and 15H02064 through Japan Society for the Promotion of Science, and World Premier International Research Center Initiative, MEXT, Japan. 

{\it Facilities:} \facility{HST (ACS, WFC3)}.

\bibliographystyle{apj}
\bibliography{reference}

\begin{thebibliography}{138}
\expandafter\ifx\csname natexlab\endcsname\relax\def\natexlab#1{#1}\fi

\bibitem[{{Ando} {et~al.}(2006){Ando}, {Ohta}, {Iwata}, {Akiyama}, {Aoki}, \&
  {Tamura}}]{2006ApJ...645L...9A}
{Ando}, M., {Ohta}, K., {Iwata}, I., {Akiyama}, M., {Aoki}, K., \& {Tamura}, N.
  2006, \apjl, 645, L9

\bibitem[{{Bacon} {et~al.}(2010){Bacon}, {Accardo}, {Adjali}, {Anwand},
  {Bauer}, {Biswas}, {Blaizot}, {Boudon}, {Brau-Nogue}, {Brinchmann},
  {Caillier}, {Capoani}, {Carollo}, {Contini}, {Couderc}, {Daguis{\'e}},
  {Deiries}, {Delabre}, {Dreizler}, {Dubois}, {Dupieux}, {Dupuy}, {Emsellem},
  {Fechner}, {Fleischmann}, {Fran{\c c}ois}, {Gallou}, {Gharsa}, {Glindemann},
  {Gojak}, {Guiderdoni}, {Hansali}, {Hahn}, {Jarno}, {Kelz}, {Koehler},
  {Kosmalski}, {Laurent}, {Le Floch}, {Lilly}, {Lizon}, {Loupias}, {Manescau},
  {Monstein}, {Nicklas}, {Olaya}, {Pares}, {Pasquini}, {P{\'e}contal-Rousset},
  {Pell{\'o}}, {Petit}, {Popow}, {Reiss}, {Remillieux}, {Renault}, {Roth},
  {Rupprecht}, {Serre}, {Schaye}, {Soucail}, {Steinmetz}, {Streicher}, {Stuik},
  {Valentin}, {Vernet}, {Weilbacher}, {Wisotzki}, \&
  {Yerle}}]{2010SPIE.7735E..08B}
{Bacon}, R., {et~al.} 2010, in \procspie, Vol. 7735, Ground-based and Airborne
  Instrumentation for Astronomy III, 773508

\bibitem[{{Bacon} {et~al.}(2015){Bacon}, {Brinchmann}, {Richard}, {Contini},
  {Drake}, {Franx}, {Tacchella}, {Vernet}, {Wisotzki}, {Blaizot}, {Bouch{\'e}},
  {Bouwens}, {Cantalupo}, {Carollo}, {Carton}, {Caruana}, {Cl{\'e}ment},
  {Dreizler}, {Epinat}, {Guiderdoni}, {Herenz}, {Husser}, {Kamann}, {Kerutt},
  {Kollatschny}, {Krajnovic}, {Lilly}, {Martinsson}, {Michel-Dansac},
  {Patricio}, {Schaye}, {Shirazi}, {Soto}, {Soucail}, {Steinmetz}, {Urrutia},
  {Weilbacher}, \& {de Zeeuw}}]{2015AA...575A..75B}
{Bacon}, R., {et~al.} 2015, \aap, 575, A75

\bibitem[{{Bacon} {et~al.}(2017){Bacon}, {Conseil}, {Mary}, {Brinchmann},
  {Shepherd}, {Akhlaghi}, {Weilbacher}, {Piqueras}, {Wisotzki}, {Lagattuta},
  {Epinat}, {Guerou}, {Inami}, {Cantalupo}, {Courbot}, {Contini}, {Richard},
  {Maseda}, {Bouwens}, {Bouch{\'e}}, {Kollatschny}, {Schaye}, {Marino},
  {Pello}, {Herenz}, {Guiderdoni}, \& {Carollo}}]{2017AA...608A...1B}
---. 2017, \aap, 608, A1

\bibitem[{{Barden} {et~al.}(2012){Barden}, {H{\"a}u{\ss}ler}, {Peng},
  {McIntosh}, \& {Guo}}]{2012MNRAS.422..449B}
{Barden}, M., {H{\"a}u{\ss}ler}, B., {Peng}, C.~Y., {McIntosh}, D.~H., \&
  {Guo}, Y. 2012, \mnras, 422, 449

\bibitem[{{Barger} {et~al.}(2012){Barger}, {Cowie}, \&
  {Wold}}]{2012ApJ...749..106B}
{Barger}, A.~J., {Cowie}, L.~L., \& {Wold}, I.~G.~B. 2012, \apj, 749, 106

\bibitem[{{Barnes} \& {Efstathiou}(1987)}]{1987ApJ...319..575B}
{Barnes}, J., \& {Efstathiou}, G. 1987, \apj, 319, 575

\bibitem[{{Beckwith} {et~al.}(2006){Beckwith}, {Stiavelli}, {Koekemoer},
  {Caldwell}, {Ferguson}, {Hook}, {Lucas}, {Bergeron}, {Corbin}, {Jogee},
  {Panagia}, {Robberto}, {Royle}, {Somerville}, \&
  {Sosey}}]{2006AJ....132.1729B}
{Beckwith}, S.~V.~W., {et~al.} 2006, \aj, 132, 1729

\bibitem[{{Behroozi} {et~al.}(2018){Behroozi}, {Wechsler}, {Hearin}, \&
  {Conroy}}]{2018arXiv180607893B}
{Behroozi}, P., {Wechsler}, R., {Hearin}, A., \& {Conroy}, C. 2018, ArXiv
  e-prints, arXiv:1806.07893

\bibitem[{{Behroozi} {et~al.}(2010){Behroozi}, {Conroy}, \&
  {Wechsler}}]{2010ApJ...717..379B}
{Behroozi}, P.~S., {Conroy}, C., \& {Wechsler}, R.~H. 2010, \apj, 717, 379

\bibitem[{{Behroozi} {et~al.}(2013){Behroozi}, {Wechsler}, \&
  {Conroy}}]{2013ApJ...770...57B}
{Behroozi}, P.~S., {Wechsler}, R.~H., \& {Conroy}, C. 2013, \apj, 770, 57

\bibitem[{{Bertin} \& {Arnouts}(1996)}]{1996A&AS..117..393B}
{Bertin}, E., \& {Arnouts}, S. 1996, \aaps, 117, 393

\bibitem[{{Bezanson} {et~al.}(2009){Bezanson}, {van Dokkum}, {Tal},
  {Marchesini}, {Kriek}, {Franx}, \& {Coppi}}]{2009ApJ...697.1290B}
{Bezanson}, R., {van Dokkum}, P.~G., {Tal}, T., {Marchesini}, D., {Kriek}, M.,
  {Franx}, M., \& {Coppi}, P. 2009, \apj, 697, 1290

\bibitem[{{Blanc} {et~al.}(2011){Blanc}, {Adams}, {Gebhardt}, {Hill}, {Drory},
  {Hao}, {Bender}, {Ciardullo}, {Finkelstein}, {Fry}, {Gawiser}, {Gronwall},
  {Hopp}, {Jeong}, {Kelzenberg}, {Komatsu}, {MacQueen}, {Murphy}, {Roth},
  {Schneider}, \& {Tufts}}]{2011ApJ...736...31B}
{Blanc}, G.~A., {et~al.} 2011, \apj, 736, 31

\bibitem[{{Bond} {et~al.}(2009){Bond}, {Gawiser}, {Gronwall}, {Ciardullo},
  {Altmann}, \& {Schawinski}}]{2009ApJ...705..639B}
{Bond}, N.~A., {Gawiser}, E., {Gronwall}, C., {Ciardullo}, R., {Altmann}, M.,
  \& {Schawinski}, K. 2009, \apj, 705, 639

\bibitem[{{Bond} {et~al.}(2012){Bond}, {Gawiser}, {Guaita}, {Padilla},
  {Gronwall}, {Ciardullo}, \& {Lai}}]{2012ApJ...753...95B}
{Bond}, N.~A., {Gawiser}, E., {Guaita}, L., {Padilla}, N., {Gronwall}, C.,
  {Ciardullo}, R., \& {Lai}, K. 2012, \apj, 753, 95

\bibitem[{{Bouwens} {et~al.}(2004){Bouwens}, {Illingworth}, {Blakeslee},
  {Broadhurst}, \& {Franx}}]{2004ApJ...611L...1B}
{Bouwens}, R.~J., {Illingworth}, G.~D., {Blakeslee}, J.~P., {Broadhurst},
  T.~J., \& {Franx}, M. 2004, \apjl, 611, L1

\bibitem[{{Bouwens} {et~al.}(2017){Bouwens}, {Illingworth}, {Oesch}, {Atek},
  {Lam}, \& {Stefanon}}]{2017ApJ...843...41B}
{Bouwens}, R.~J., {Illingworth}, G.~D., {Oesch}, P.~A., {Atek}, H., {Lam}, D.,
  \& {Stefanon}, M. 2017, \apj, 843, 41

\bibitem[{{Bouwens} {et~al.}(2011){Bouwens}, {Illingworth}, {Oesch},
  {Labb{\'e}}, {Trenti}, {van Dokkum}, {Franx}, {Stiavelli}, {Carollo},
  {Magee}, \& {Gonzalez}}]{2011ApJ...737...90B}
{Bouwens}, R.~J., {et~al.} 2011, \apj, 737, 90

\bibitem[{{Brooks} {et~al.}(2009){Brooks}, {Governato}, {Quinn}, {Brook}, \&
  {Wadsley}}]{2009ApJ...694..396B}
{Brooks}, A.~M., {Governato}, F., {Quinn}, T., {Brook}, C.~B., \& {Wadsley}, J.
  2009, \apj, 694, 396

\bibitem[{{Bryan} \& {Norman}(1998)}]{1998ApJ...495...80B}
{Bryan}, G.~L., \& {Norman}, M.~L. 1998, \apj, 495, 80

\bibitem[{{Bullock} {et~al.}(2001){Bullock}, {Dekel}, {Kolatt}, {Kravtsov},
  {Klypin}, {Porciani}, \& {Primack}}]{2001ApJ...555..240B}
{Bullock}, J.~S., {Dekel}, A., {Kolatt}, T.~S., {Kravtsov}, A.~V., {Klypin},
  A.~A., {Porciani}, C., \& {Primack}, J.~R. 2001, \apj, 555, 240

\bibitem[{{Caruana} {et~al.}(2018){Caruana}, {Wisotzki}, {Herenz}, {Kerutt},
  {Urrutia}, {Schmidt}, {Bouwens}, {Brinchmann}, {Cantalupo}, {Carollo},
  {Diener}, {Drake}, {Garel}, {Marino}, {Richard}, {Saust}, {Schaye}, \&
  {Verhamme}}]{2018MNRAS.473...30C}
{Caruana}, J., {et~al.} 2018, \mnras, 473, 30

\bibitem[{{Ciardullo} {et~al.}(2012){Ciardullo}, {Gronwall}, {Wolf},
  {McCathran}, {Bond}, {Gawiser}, {Guaita}, {Feldmeier}, {Treister}, {Padilla},
  {Francke}, {Matkovi{\'c}}, {Altmann}, \& {Herrera}}]{2012ApJ...744..110C}
{Ciardullo}, R., {et~al.} 2012, \apj, 744, 110

\bibitem[{{Cowie} {et~al.}(2010){Cowie}, {Barger}, \&
  {Hu}}]{2010ApJ...711..928C}
{Cowie}, L.~L., {Barger}, A.~J., \& {Hu}, E.~M. 2010, \apj, 711, 928

\bibitem[{{Cowie} {et~al.}(2011){Cowie}, {Hu}, \&
  {Songaila}}]{2011ApJ...735L..38C}
{Cowie}, L.~L., {Hu}, E.~M., \& {Songaila}, A. 2011, \apjl, 735, L38

\bibitem[{{Davis} \& {Natarajan}(2009)}]{2009MNRAS.393.1498D}
{Davis}, A.~J., \& {Natarajan}, P. 2009, \mnras, 393, 1498

\bibitem[{{Dayal} \& {Ferrara}(2018)}]{2018arXiv180909136D}
{Dayal}, P., \& {Ferrara}, A. 2018, ArXiv e-prints, arXiv:1809.09136

\bibitem[{{Deharveng} {et~al.}(2008){Deharveng}, {Small}, {Barlow},
  {P{\'e}roux}, {Milliard}, {Friedman}, {Martin}, {Morrissey}, {Schiminovich},
  {Forster}, {Seibert}, {Wyder}, {Bianchi}, {Donas}, {Heckman}, {Lee},
  {Madore}, {Neff}, {Rich}, {Szalay}, {Welsh}, \& {Yi}}]{2008ApJ...680.1072D}
{Deharveng}, J.-M., {et~al.} 2008, \apj, 680, 1072

\bibitem[{{Dijkstra}(2017)}]{2017arXiv170403416D}
{Dijkstra}, M. 2017, ArXiv e-prints, arXiv:1704.03416

\bibitem[{{Dijkstra} \& {Kramer}(2012)}]{2012MNRAS.424.1672D}
{Dijkstra}, M., \& {Kramer}, R. 2012, \mnras, 424, 1672

\bibitem[{{Ellis} {et~al.}(2013){Ellis}, {McLure}, {Dunlop}, {Robertson},
  {Ono}, {Schenker}, {Koekemoer}, {Bowler}, {Ouchi}, {Rogers}, {Curtis-Lake},
  {Schneider}, {Charlot}, {Stark}, {Furlanetto}, \&
  {Cirasuolo}}]{2013ApJ...763L...7E}
{Ellis}, R.~S., {et~al.} 2013, \apjl, 763, L7

\bibitem[{{Fall} \& {Efstathiou}(1980)}]{1980MNRAS.193..189F}
{Fall}, S.~M., \& {Efstathiou}, G. 1980, \mnras, 193, 189

\bibitem[{{Fall} \& {Romanowsky}(2018)}]{2018arXiv180802525F}
{Fall}, S.~M., \& {Romanowsky}, A.~J. 2018, ArXiv e-prints

\bibitem[{{Feldmeier} {et~al.}(2013){Feldmeier}, {Hagen}, {Ciardullo},
  {Gronwall}, {Gawiser}, {Guaita}, {Hagen}, {Bond}, {Acquaviva}, {Blanc},
  {Orsi}, \& {Kurczynski}}]{2013ApJ...776...75F}
{Feldmeier}, J.~J., {et~al.} 2013, \apj, 776, 75

\bibitem[{{Ferguson} {et~al.}(2004){Ferguson}, {Dickinson}, {Giavalisco},
  {Kretchmer}, {Ravindranath}, {Idzi}, {Taylor}, {Conselice}, {Fall},
  {Gardner}, {Livio}, {Madau}, {Moustakas}, {Papovich}, {Somerville},
  {Spinrad}, \& {Stern}}]{2004ApJ...600L.107F}
{Ferguson}, H.~C., {et~al.} 2004, \apjl, 600, L107

\bibitem[{{Finkelstein} {et~al.}(2009){Finkelstein}, {Rhoads}, {Malhotra}, \&
  {Grogin}}]{2009ApJ...691..465F}
{Finkelstein}, S.~L., {Rhoads}, J.~E., {Malhotra}, S., \& {Grogin}, N. 2009,
  \apj, 691, 465

\bibitem[{{Grazian} {et~al.}(2012){Grazian}, {Castellano}, {Fontana},
  {Pentericci}, {Dunlop}, {McLure}, {Koekemoer}, {Dickinson}, {Faber},
  {Ferguson}, {Galametz}, {Giavalisco}, {Grogin}, {Hathi}, {Kocevski}, {Lai},
  {Newman}, \& {Vanzella}}]{2012AA...547A..51G}
{Grazian}, A., {et~al.} 2012, \aap, 547, A51

\bibitem[{{Grogin} {et~al.}(2011){Grogin}, {Kocevski}, {Faber}, {Ferguson},
  {Koekemoer}, {Riess}, {Acquaviva}, {Alexander}, {Almaini}, {Ashby}, {Barden},
  {Bell}, {Bournaud}, {Brown}, {Caputi}, {Casertano}, {Cassata}, {Castellano},
  {Challis}, {Chary}, {Cheung}, {Cirasuolo}, {Conselice}, {Roshan Cooray},
  {Croton}, {Daddi}, {Dahlen}, {Dav{\'e}}, {de Mello}, {Dekel}, {Dickinson},
  {Dolch}, {Donley}, {Dunlop}, {Dutton}, {Elbaz}, {Fazio}, {Filippenko},
  {Finkelstein}, {Fontana}, {Gardner}, {Garnavich}, {Gawiser}, {Giavalisco},
  {Grazian}, {Guo}, {Hathi}, {H{\"a}ussler}, {Hopkins}, {Huang}, {Huang},
  {Jha}, {Kartaltepe}, {Kirshner}, {Koo}, {Lai}, {Lee}, {Li}, {Lotz}, {Lucas},
  {Madau}, {McCarthy}, {McGrath}, {McIntosh}, {McLure}, {Mobasher},
  {Moustakas}, {Mozena}, {Nandra}, {Newman}, {Niemi}, {Noeske}, {Papovich},
  {Pentericci}, {Pope}, {Primack}, {Rajan}, {Ravindranath}, {Reddy}, {Renzini},
  {Rix}, {Robaina}, {Rodney}, {Rosario}, {Rosati}, {Salimbeni}, {Scarlata},
  {Siana}, {Simard}, {Smidt}, {Somerville}, {Spinrad}, {Straughn}, {Strolger},
  {Telford}, {Teplitz}, {Trump}, {van der Wel}, {Villforth}, {Wechsler},
  {Weiner}, {Wiklind}, {Wild}, {Wilson}, {Wuyts}, {Yan}, \&
  {Yun}}]{2011ApJS..197...35G}
{Grogin}, N.~A., {et~al.} 2011, \apjs, 197, 35

\bibitem[{{Gronwall} {et~al.}(2011){Gronwall}, {Bond}, {Ciardullo}, {Gawiser},
  {Altmann}, {Blanc}, \& {Feldmeier}}]{2011ApJ...743....9G}
{Gronwall}, C., {Bond}, N.~A., {Ciardullo}, R., {Gawiser}, E., {Altmann}, M.,
  {Blanc}, G.~A., \& {Feldmeier}, J.~J. 2011, \apj, 743, 9

\bibitem[{{Gronwall} {et~al.}(2007){Gronwall}, {Ciardullo}, {Hickey},
  {Gawiser}, {Feldmeier}, {van Dokkum}, {Urry}, {Herrera}, {Lehmer}, {Infante},
  {Orsi}, {Marchesini}, {Blanc}, {Francke}, {Lira}, \&
  {Treister}}]{2007ApJ...667...79G}
{Gronwall}, C., {et~al.} 2007, \apj, 667, 79

\bibitem[{{Guaita} {et~al.}(2015){Guaita}, {Melinder}, {Hayes}, {{\"O}stlin},
  {Gonzalez}, {Micheva}, {Adamo}, {Mas-Hesse}, {Sandberg},
  {Ot{\'{\i}}-Floranes}, {Schaerer}, {Verhamme}, {Freeland}, {Orlitov{\'a}},
  {Laursen}, {Cannon}, {Duval}, {Rivera-Thorsen}, {Herenz}, {Kunth}, {Atek},
  {Puschnig}, {Gruyters}, \& {Pardy}}]{2015AA...576A..51G}
{Guaita}, L., {et~al.} 2015, \aap, 576, A51

\bibitem[{{Hagen} {et~al.}(2014){Hagen}, {Ciardullo}, {Gronwall}, {Acquaviva},
  {Bridge}, {Zeimann}, {Blanc}, {Bond}, {Finkelstein}, {Song}, {Gawiser},
  {Fox}, {Gebhardt}, {Malz}, {Schneider}, {Drory}, {Gebhardt}, \&
  {Hill}}]{2014ApJ...786...59H}
{Hagen}, A., {et~al.} 2014, \apj, 786, 59

\bibitem[{{Hagen} {et~al.}(2016){Hagen}, {Zeimann}, {Behrens}, {Ciardullo},
  {Grasshorn Gebhardt}, {Gronwall}, {Bridge}, {Fox}, {Schneider}, {Trump},
  {Blanc}, {Chiang}, {Chonis}, {Finkelstein}, {Hill}, {Jogee}, \&
  {Gawiser}}]{2016ApJ...817...79H}
---. 2016, \apj, 817, 79

\bibitem[{{Harikane} {et~al.}(2016){Harikane}, {Ouchi}, {Ono}, {More}, {Saito},
  {Lin}, {Coupon}, {Shimasaku}, {Shibuya}, {Price}, {Lin}, {Hsieh}, {Ishigaki},
  {Komiyama}, {Silverman}, {Takata}, {Tamazawa}, \&
  {Toshikawa}}]{2016ApJ...821..123H}
{Harikane}, Y., {et~al.} 2016, \apj, 821, 123

\bibitem[{{Hashimoto} {et~al.}(2015){Hashimoto}, {Verhamme}, {Ouchi},
  {Shimasaku}, {Schaerer}, {Nakajima}, {Shibuya}, {Rauch}, {Ono}, \&
  {Goto}}]{2015ApJ...812..157H}
{Hashimoto}, T., {et~al.} 2015, \apj, 812, 157

\bibitem[{{Hayashino} {et~al.}(2004){Hayashino}, {Matsuda}, {Tamura},
  {Yamauchi}, {Yamada}, {Ajiki}, {Fujita}, {Murayama}, {Nagao}, {Ohta},
  {Okamura}, {Ouchi}, {Shimasaku}, {Shioya}, \&
  {Taniguchi}}]{2004AJ....128.2073H}
{Hayashino}, T., {et~al.} 2004, \aj, 128, 2073

\bibitem[{{Hayes} {et~al.}(2014){Hayes}, {{\"O}stlin}, {Duval}, {Sandberg},
  {Guaita}, {Melinder}, {Adamo}, {Schaerer}, {Verhamme}, {Orlitov{\'a}},
  {Mas-Hesse}, {Cannon}, {Atek}, {Kunth}, {Laursen}, {Ot{\'{\i}}-Floranes},
  {Pardy}, {Rivera-Thorsen}, \& {Herenz}}]{2014ApJ...782....6H}
{Hayes}, M., {et~al.} 2014, \apj, 782, 6

\bibitem[{{Hearin} {et~al.}(2017){Hearin}, {Behroozi}, {Kravtsov}, \&
  {Moster}}]{2017arXiv171110500H}
{Hearin}, A., {Behroozi}, P., {Kravtsov}, A., \& {Moster}, B. 2017, ArXiv
  e-prints

\bibitem[{{Herenz} {et~al.}(2016){Herenz}, {Gruyters}, {Orlitova}, {Hayes},
  {{\"O}stlin}, {Cannon}, {Roth}, {Bik}, {Pardy}, {Ot{\'{\i}}-Floranes},
  {Mas-Hesse}, {Adamo}, {Atek}, {Duval}, {Guaita}, {Kunth}, {Laursen},
  {Melinder}, {Puschnig}, {Rivera-Thorsen}, {Schaerer}, \&
  {Verhamme}}]{2016A&A...587A..78H}
{Herenz}, E.~C., {et~al.} 2016, \aap, 587, A78

\bibitem[{{Herenz} {et~al.}(2017){Herenz}, {Urrutia}, {Wisotzki}, {Kerutt},
  {Saust}, {Werhahn}, {Schmidt}, {Caruana}, {Diener}, {Bacon}, {Brinchmann},
  {Schaye}, {Maseda}, \& {Weilbacher}}]{2017A&A...606A..12H}
---. 2017, \aap, 606, A12

\bibitem[{{Holwerda} {et~al.}(2015){Holwerda}, {Bouwens}, {Oesch}, {Smit},
  {Illingworth}, \& {Labbe}}]{2015ApJ...808....6H}
{Holwerda}, B.~W., {Bouwens}, R., {Oesch}, P., {Smit}, R., {Illingworth}, G.,
  \& {Labbe}, I. 2015, \apj, 808, 6

\bibitem[{{Hu} {et~al.}(2010){Hu}, {Cowie}, {Barger}, {Capak}, {Kakazu}, \&
  {Trouille}}]{2010ApJ...725..394H}
{Hu}, E.~M., {Cowie}, L.~L., {Barger}, A.~J., {Capak}, P., {Kakazu}, Y., \&
  {Trouille}, L. 2010, \apj, 725, 394

\bibitem[{{Hu} {et~al.}(2017){Hu}, {Wang}, {Zheng}, {Malhotra}, {Infante},
  {Rhoads}, {Gonzalez}, {Walker}, {Jiang}, {Jiang}, {Hibon}, {Barrientos},
  {Finkelstein}, {Galaz}, {Kang}, {Kong}, {Tilvi}, {Yang}, \&
  {Zheng}}]{2017ApJ...845L..16H}
{Hu}, W., {et~al.} 2017, \apj, 845, L16

\bibitem[{{Huang} {et~al.}(2013){Huang}, {Ferguson}, {Ravindranath}, \&
  {Su}}]{2013ApJ...765...68H}
{Huang}, K.-H., {Ferguson}, H.~C., {Ravindranath}, S., \& {Su}, J. 2013, \apj,
  765, 68

\bibitem[{{Huang} {et~al.}(2017){Huang}, {Fall}, {Ferguson}, {van der Wel},
  {Grogin}, {Koekemoer}, {Lee}, {P{\'e}rez-Gonz{\'a}lez}, \&
  {Wuyts}}]{2017ApJ...838....6H}
{Huang}, K.-H., {et~al.} 2017, \apj, 838, 6

\bibitem[{{Illingworth} {et~al.}(2013){Illingworth}, {Magee}, {Oesch},
  {Bouwens}, {Labb{\'e}}, {Stiavelli}, {van Dokkum}, {Franx}, {Trenti},
  {Carollo}, \& {Gonzalez}}]{2013ApJS..209....6I}
{Illingworth}, G.~D., {et~al.} 2013, \apjs, 209, 6

\bibitem[{{Itoh} {et~al.}(2018){Itoh}, {Ouchi}, {Zhang}, {Inoue}, {Mawatari},
  {Shibuya}, {Harikane}, {Ono}, {Kusakabe}, {Shimasaku}, {Fujimoto}, {Iwata},
  {Kajisawa}, {Kashikawa}, {Kawanomoto}, {Komiyama}, {Lee}, {Nagao}, \&
  {Taniguchi}}]{2018arXiv180505944I}
{Itoh}, R., {et~al.} 2018, ArXiv e-prints, arXiv:1805.05944

\bibitem[{{Jiang} {et~al.}(2013){Jiang}, {Egami}, {Fan}, {Windhorst}, {Cohen},
  {Dav{\'e}}, {Finlator}, {Kashikawa}, {Mechtley}, {Ouchi}, \&
  {Shimasaku}}]{2013ApJ...773..153J}
{Jiang}, L., {et~al.} 2013, \apj, 773, 153

\bibitem[{{Jiang} {et~al.}(2016){Jiang}, {Finlator}, {Cohen}, {Egami},
  {Windhorst}, {Fan}, {Dav{\'e}}, {Kashikawa}, {Mechtley}, {Ouchi},
  {Shimasaku}, \& {Cl{\'e}ment}}]{2016ApJ...816...16J}
---. 2016, \apj, 816, 16

\bibitem[{{Kawamata} {et~al.}(2015){Kawamata}, {Ishigaki}, {Shimasaku},
  {Oguri}, \& {Ouchi}}]{2015ApJ...804..103K}
{Kawamata}, R., {Ishigaki}, M., {Shimasaku}, K., {Oguri}, M., \& {Ouchi}, M.
  2015, \apj, 804, 103

\bibitem[{{Kawamata} {et~al.}(2018){Kawamata}, {Ishigaki}, {Shimasaku},
  {Oguri}, {Ouchi}, \& {Tanigawa}}]{2018ApJ...855....4K}
{Kawamata}, R., {Ishigaki}, M., {Shimasaku}, K., {Oguri}, M., {Ouchi}, M., \&
  {Tanigawa}, S. 2018, \apj, 855, 4

\bibitem[{{Kelvin} {et~al.}(2012){Kelvin}, {Driver}, {Robotham}, {Hill},
  {Alpaslan}, {Baldry}, {Bamford}, {Bland-Hawthorn}, {Brough}, {Graham},
  {H{\"a}ussler}, {Hopkins}, {Liske}, {Loveday}, {Norberg}, {Phillipps},
  {Popescu}, {Prescott}, {Taylor}, \& {Tuffs}}]{2012MNRAS.421.1007K}
{Kelvin}, L.~S., {et~al.} 2012, \mnras, 421, 1007

\bibitem[{{Kennicutt}(1998)}]{1998ARA&A..36..189K}
{Kennicutt}, Jr., R.~C. 1998, \araa, 36, 189

\bibitem[{{Kobayashi} {et~al.}(2016){Kobayashi}, {Murata}, {Koekemoer},
  {Murayama}, {Taniguchi}, {Kajisawa}, {Shioya}, {Scoville}, {Nagao}, \&
  {Capak}}]{2016ApJ...819...25K}
{Kobayashi}, M.~A.~R., {et~al.} 2016, \apj, 819, 25

\bibitem[{{Koekemoer} {et~al.}(2011){Koekemoer}, {Faber}, {Ferguson}, {Grogin},
  {Kocevski}, {Koo}, {Lai}, {Lotz}, {Lucas}, {McGrath}, {Ogaz}, {Rajan},
  {Riess}, {Rodney}, {Strolger}, {Casertano}, {Castellano}, {Dahlen},
  {Dickinson}, {Dolch}, {Fontana}, {Giavalisco}, {Grazian}, {Guo}, {Hathi},
  {Huang}, {van der Wel}, {Yan}, {Acquaviva}, {Alexander}, {Almaini}, {Ashby},
  {Barden}, {Bell}, {Bournaud}, {Brown}, {Caputi}, {Cassata}, {Challis},
  {Chary}, {Cheung}, {Cirasuolo}, {Conselice}, {Roshan Cooray}, {Croton},
  {Daddi}, {Dav{\'e}}, {de Mello}, {de Ravel}, {Dekel}, {Donley}, {Dunlop},
  {Dutton}, {Elbaz}, {Fazio}, {Filippenko}, {Finkelstein}, {Frazer}, {Gardner},
  {Garnavich}, {Gawiser}, {Gruetzbauch}, {Hartley}, {H{\"a}ussler},
  {Herrington}, {Hopkins}, {Huang}, {Jha}, {Johnson}, {Kartaltepe},
  {Khostovan}, {Kirshner}, {Lani}, {Lee}, {Li}, {Madau}, {McCarthy},
  {McIntosh}, {McLure}, {McPartland}, {Mobasher}, {Moreira}, {Mortlock},
  {Moustakas}, {Mozena}, {Nandra}, {Newman}, {Nielsen}, {Niemi}, {Noeske},
  {Papovich}, {Pentericci}, {Pope}, {Primack}, {Ravindranath}, {Reddy},
  {Renzini}, {Rix}, {Robaina}, {Rosario}, {Rosati}, {Salimbeni}, {Scarlata},
  {Siana}, {Simard}, {Smidt}, {Snyder}, {Somerville}, {Spinrad}, {Straughn},
  {Telford}, {Teplitz}, {Trump}, {Vargas}, {Villforth}, {Wagner}, {Wandro},
  {Wechsler}, {Weiner}, {Wiklind}, {Wild}, {Wilson}, {Wuyts}, \&
  {Yun}}]{2011ApJS..197...36K}
{Koekemoer}, A.~M., {et~al.} 2011, \apjs, 197, 36

\bibitem[{{Konno} {et~al.}(2016){Konno}, {Ouchi}, {Nakajima}, {Duval},
  {Kusakabe}, {Ono}, \& {Shimasaku}}]{2016ApJ...823...20K}
{Konno}, A., {Ouchi}, M., {Nakajima}, K., {Duval}, F., {Kusakabe}, H., {Ono},
  Y., \& {Shimasaku}, K. 2016, \apj, 823, 20

\bibitem[{{Lake} {et~al.}(2015){Lake}, {Zheng}, {Cen}, {Sadoun}, {Momose}, \&
  {Ouchi}}]{2015ApJ...806...46L}
{Lake}, E., {Zheng}, Z., {Cen}, R., {Sadoun}, R., {Momose}, R., \& {Ouchi}, M.
  2015, \apj, 806, 46

\bibitem[{{Leclercq} {et~al.}(2017){Leclercq}, {Bacon}, {Wisotzki}, {Mitchell},
  {Garel}, {Verhamme}, {Blaizot}, {Hashimoto}, {Herenz}, {Conseil},
  {Cantalupo}, {Inami}, {Contini}, {Richard}, {Maseda}, {Schaye}, {Marino},
  {Akhlaghi}, {Brinchmann}, \& {Carollo}}]{2017AA...608A...8L}
{Leclercq}, F., {et~al.} 2017, \aap, 608, A8

\bibitem[{{Malhotra} {et~al.}(2012){Malhotra}, {Rhoads}, {Finkelstein},
  {Hathi}, {Nilsson}, {McLinden}, \& {Pirzkal}}]{2012ApJ...750L..36M}
{Malhotra}, S., {Rhoads}, J.~E., {Finkelstein}, S.~L., {Hathi}, N., {Nilsson},
  K., {McLinden}, E., \& {Pirzkal}, N. 2012, \apjl, 750, L36

\bibitem[{{Mas-Ribas} \& {Dijkstra}(2016)}]{2016ApJ...822...84M}
{Mas-Ribas}, L., \& {Dijkstra}, M. 2016, \apj, 822, 84

\bibitem[{{Mas-Ribas} {et~al.}(2017{\natexlab{a}}){Mas-Ribas}, {Dijkstra},
  {Hennawi}, {Trenti}, {Momose}, \& {Ouchi}}]{2017ApJ...841...19M}
{Mas-Ribas}, L., {Dijkstra}, M., {Hennawi}, J.~F., {Trenti}, M., {Momose}, R.,
  \& {Ouchi}, M. 2017{\natexlab{a}}, \apj, 841, 19

\bibitem[{{Mas-Ribas} {et~al.}(2017{\natexlab{b}}){Mas-Ribas}, {Hennawi},
  {Dijkstra}, {Davies}, {Stern}, \& {Rix}}]{2017ApJ...846...11M}
{Mas-Ribas}, L., {Hennawi}, J.~F., {Dijkstra}, M., {Davies}, F.~B., {Stern},
  J., \& {Rix}, H.-W. 2017{\natexlab{b}}, \apj, 846, 11

\bibitem[{{Matsuda} {et~al.}(2012){Matsuda}, {Yamada}, {Hayashino}, {Yamauchi},
  {Nakamura}, {Morimoto}, {Ouchi}, {Ono}, {Umemura}, \&
  {Mori}}]{2012MNRAS.425..878M}
{Matsuda}, Y., {et~al.} 2012, \mnras, 425, 878

\bibitem[{{Matthee} {et~al.}(2016){Matthee}, {Sobral}, {Oteo}, {Best}, {Smail},
  {R{\"o}ttgering}, \& {Paulino-Afonso}}]{2016MNRAS.458..449M}
{Matthee}, J., {Sobral}, D., {Oteo}, I., {Best}, P., {Smail}, I.,
  {R{\"o}ttgering}, H., \& {Paulino-Afonso}, A. 2016, \mnras, 458, 449

\bibitem[{{Matthee} {et~al.}(2015){Matthee}, {Sobral}, {Santos},
  {R{\"o}ttgering}, {Darvish}, \& {Mobasher}}]{2015MNRAS.451..400M}
{Matthee}, J., {Sobral}, D., {Santos}, S., {R{\"o}ttgering}, H., {Darvish}, B.,
  \& {Mobasher}, B. 2015, \mnras, 451, 400

\bibitem[{{McLure} {et~al.}(2013){McLure}, {Pearce}, {Dunlop}, {Cirasuolo},
  {Curtis-Lake}, {Bruce}, {Caputi}, {Almaini}, {Bonfield}, {Bradshaw},
  {Buitrago}, {Chuter}, {Foucaud}, {Hartley}, \&
  {Jarvis}}]{2013MNRAS.428.1088M}
{McLure}, R.~J., {et~al.} 2013, \mnras, 428, 1088

\bibitem[{{Mo} {et~al.}(1998){Mo}, {Mao}, \& {White}}]{1998MNRAS.295..319M}
{Mo}, H.~J., {Mao}, S., \& {White}, S.~D.~M. 1998, \mnras, 295, 319

\bibitem[{{Momose} {et~al.}(2014){Momose}, {Ouchi}, {Nakajima}, {Ono},
  {Shibuya}, {Shimasaku}, {Yuma}, {Mori}, \& {Umemura}}]{2014MNRAS.442..110M}
{Momose}, R., {et~al.} 2014, \mnras, 442, 110

\bibitem[{{Momose} {et~al.}(2016){Momose}, {Ouchi}, {Nakajima}, {Ono},
  {Shibuya}, {Shimasaku}, {Yuma}, {Mori}, \& {Umemura}}]{2016MNRAS.457.2318M}
---. 2016, \mnras, 457, 2318

\bibitem[{{Mosleh} {et~al.}(2013){Mosleh}, {Williams}, \&
  {Franx}}]{2013ApJ...777..117M}
{Mosleh}, M., {Williams}, R.~J., \& {Franx}, M. 2013, \apj, 777, 117

\bibitem[{{Mosleh} {et~al.}(2012){Mosleh}, {Williams}, {Franx}, {Gonzalez},
  {Bouwens}, {Oesch}, {Labbe}, {Illingworth}, \&
  {Trenti}}]{2012ApJ...756L..12M}
{Mosleh}, M., {et~al.} 2012, \apjl, 756, L12

\bibitem[{{Murayama} {et~al.}(2007){Murayama}, {Taniguchi}, {Scoville},
  {Ajiki}, {Sanders}, {Mobasher}, {Aussel}, {Capak}, {Koekemoer}, {Shioya},
  {Nagao}, {Carilli}, {Ellis}, {Garilli}, {Giavalisco}, {Kitzbichler}, {Le
  F{\`e}vre}, {Maccagni}, {Schinnerer}, {Smol{\v c}i{\'c}}, {Tribiano},
  {Cimatti}, {Komiyama}, {Miyazaki}, {Sasaki}, {Koda}, \&
  {Karoji}}]{2007ApJS..172..523M}
{Murayama}, T., {et~al.} 2007, \apjs, 172, 523

\bibitem[{{Nakajima} \& {Ouchi}(2013)}]{2013arXiv1309.0207N}
{Nakajima}, K., \& {Ouchi}, M. 2013, ArXiv e-prints

\bibitem[{{Nakajima} {et~al.}(2012){Nakajima}, {Ouchi}, {Shimasaku}, {Ono},
  {Lee}, {Foucaud}, {Ly}, {Dale}, {Salim}, {Finn}, {Almaini}, \&
  {Okamura}}]{2012ApJ...745...12N}
{Nakajima}, K., {et~al.} 2012, \apj, 745, 12

\bibitem[{{Oesch} {et~al.}(2015){Oesch}, {van Dokkum}, {Illingworth},
  {Bouwens}, {Momcheva}, {Holden}, {Roberts-Borsani}, {Smit}, {Franx},
  {Labb{\'e}}, {Gonz{\'a}lez}, \& {Magee}}]{2015ApJ...804L..30O}
{Oesch}, P.~A., {et~al.} 2015, \apj, 804, L30

\bibitem[{{Okamura} {et~al.}(2018){Okamura}, {Shimasaku}, \&
  {Kawamata}}]{2018ApJ...854...22O}
{Okamura}, T., {Shimasaku}, K., \& {Kawamata}, R. 2018, \apj, 854, 22

\bibitem[{{Oke} \& {Gunn}(1983)}]{1983ApJ...266..713O}
{Oke}, J.~B., \& {Gunn}, J.~E. 1983, \apj, 266, 713

\bibitem[{{Ono} {et~al.}(2012){Ono}, {Ouchi}, {Mobasher}, {Dickinson},
  {Penner}, {Shimasaku}, {Weiner}, {Kartaltepe}, {Nakajima}, {Nayyeri},
  {Stern}, {Kashikawa}, \& {Spinrad}}]{2012ApJ...744...83O}
{Ono}, Y., {et~al.} 2012, \apj, 744, 83

\bibitem[{{Ono} {et~al.}(2013){Ono}, {Ouchi}, {Curtis-Lake}, {Schenker},
  {Ellis}, {McLure}, {Dunlop}, {Robertson}, {Koekemoer}, {Bowler}, {Rogers},
  {Schneider}, {Charlot}, {Stark}, {Shimasaku}, {Furlanetto}, \&
  {Cirasuolo}}]{2013ApJ...777..155O}
---. 2013, \apj, 777, 155

\bibitem[{{{\"O}stlin} {et~al.}(2014){{\"O}stlin}, {Hayes}, {Duval},
  {Sandberg}, {Rivera-Thorsen}, {Marquart}, {Orlitov{\'a}}, {Adamo},
  {Melinder}, {Guaita}, {Atek}, {Cannon}, {Gruyters}, {Herenz}, {Kunth},
  {Laursen}, {Mas-Hesse}, {Micheva}, {Ot{\'{\i}}-Floranes}, {Pardy}, {Roth},
  {Schaerer}, \& {Verhamme}}]{2014ApJ...797...11O}
{{\"O}stlin}, G., {et~al.} 2014, \apj, 797, 11

\bibitem[{{Ota} {et~al.}(2017){Ota}, {Iye}, {Kashikawa}, {Konno}, {Nakata},
  {Totani}, {Kobayashi}, {Fudamoto}, {Seko}, {Toshikawa}, {Ichikawa},
  {Shibuya}, \& {Onoue}}]{2017ApJ...844...85O}
{Ota}, K., {et~al.} 2017, \apj, 844, 85

\bibitem[{{Ouchi} {et~al.}(2008){Ouchi}, {Shimasaku}, {Akiyama}, {Simpson},
  {Saito}, {Ueda}, {Furusawa}, {Sekiguchi}, {Yamada}, {Kodama}, {Kashikawa},
  {Okamura}, {Iye}, {Takata}, {Yoshida}, \& {Yoshida}}]{2008ApJS..176..301O}
{Ouchi}, M., {et~al.} 2008, \apjs, 176, 301

\bibitem[{{Ouchi} {et~al.}(2009){Ouchi}, {Mobasher}, {Shimasaku}, {Ferguson},
  {Fall}, {Ono}, {Kashikawa}, {Morokuma}, {Nakajima}, {Okamura}, {Dickinson},
  {Giavalisco}, \& {Ohta}}]{2009ApJ...706.1136O}
---. 2009, \apj, 706, 1136

\bibitem[{{Ouchi} {et~al.}(2010){Ouchi}, {Shimasaku}, {Furusawa}, {Saito},
  {Yoshida}, {Akiyama}, {Ono}, {Yamada}, {Ota}, {Kashikawa}, {Iye}, {Kodama},
  {Okamura}, {Simpson}, \& {Yoshida}}]{2010ApJ...723..869O}
---. 2010, \apj, 723, 869

\bibitem[{{Overzier} {et~al.}(2008){Overzier}, {Bouwens}, {Cross}, {Venemans},
  {Miley}, {Zirm}, {Ben{\'{\i}}tez}, {Blakeslee}, {Coe}, {Demarco}, {Ford},
  {Homeier}, {Illingworth}, {Kurk}, {Martel}, {Mei}, {Oliveira},
  {R{\"o}ttgering}, {Tsvetanov}, \& {Zheng}}]{2008ApJ...673..143O}
{Overzier}, R.~A., {et~al.} 2008, \apj, 673, 143

\bibitem[{{Pardy} {et~al.}(2014){Pardy}, {Cannon}, {{\"O}stlin}, {Hayes},
  {Rivera-Thorsen}, {Sandberg}, {Adamo}, {Freeland}, {Herenz}, {Guaita},
  {Kunth}, {Laursen}, {Mas-Hesse}, {Melinder}, {Orlitov{\'a}},
  {Ot{\'\i}-Floranes}, {Puschnig}, {Schaerer}, \&
  {Verhamme}}]{2014ApJ...794..101P}
{Pardy}, S.~A., {et~al.} 2014, \apj, 794, 101

\bibitem[{{Paulino-Afonso} {et~al.}(2017){Paulino-Afonso}, {Sobral},
  {Buitrago}, \& {Afonso}}]{2017MNRAS.465.2717P}
{Paulino-Afonso}, A., {Sobral}, D., {Buitrago}, F., \& {Afonso}, J. 2017,
  \mnras, 465, 2717

\bibitem[{{Paulino-Afonso} {et~al.}(2018){Paulino-Afonso}, {Sobral}, {Ribeiro},
  {Matthee}, {Santos}, {Calhau}, {Forshaw}, {Johnson}, {Merrick}, {P{\'e}rez},
  \& {Sheldon}}]{2018MNRAS.476.5479P}
{Paulino-Afonso}, A., {et~al.} 2018, \mnras, 476, 5479

\bibitem[{{Peng} {et~al.}(2002){Peng}, {Ho}, {Impey}, \&
  {Rix}}]{2002AJ....124..266P}
{Peng}, C.~Y., {Ho}, L.~C., {Impey}, C.~D., \& {Rix}, H.-W. 2002, \aj, 124, 266

\bibitem[{{Peng} {et~al.}(2010){Peng}, {Ho}, {Impey}, \&
  {Rix}}]{2010AJ....139.2097P}
---. 2010, \aj, 139, 2097

\bibitem[{{Pentericci} {et~al.}(2018){Pentericci}, {Vanzella}, {Castellano},
  {Fontana}, {De Barros}, {Grazian}, {Marchi}, {Bradac}, {Conselice},
  {Cristiani}, {Dickinson}, {Finkelstein}, {Giallongo}, {Guaita}, {Koekemoer},
  {Maiolino}, {Santini}, \& {Tilvi}}]{2018arXiv180801847P}
{Pentericci}, L., {et~al.} 2018, ArXiv e-prints, arXiv:1808.01847

\bibitem[{{Pirzkal} {et~al.}(2007){Pirzkal}, {Malhotra}, {Rhoads}, \&
  {Xu}}]{2007ApJ...667...49P}
{Pirzkal}, N., {Malhotra}, S., {Rhoads}, J.~E., \& {Xu}, C. 2007, \apj, 667, 49

\bibitem[{{Planck Collaboration} {et~al.}(2016){Planck Collaboration}, {Ade},
  {Aghanim}, {Arnaud}, {Ashdown}, {Aumont}, {Baccigalupi}, {Banday},
  {Barreiro}, {Bartlett}, \& et~al.}]{2016AA...594A..13P}
{Planck Collaboration} {et~al.} 2016, \aap, 594, A13

\bibitem[{{Prada} {et~al.}(2012){Prada}, {Klypin}, {Cuesta}, {Betancort-Rijo},
  \& {Primack}}]{2012MNRAS.423.3018P}
{Prada}, F., {Klypin}, A.~A., {Cuesta}, A.~J., {Betancort-Rijo}, J.~E., \&
  {Primack}, J. 2012, \mnras, 423, 3018

\bibitem[{{Rauch} {et~al.}(2008){Rauch}, {Haehnelt}, {Bunker}, {Becker},
  {Marleau}, {Graham}, {Cristiani}, {Jarvis}, {Lacey}, {Morris}, {Peroux},
  {R{\"o}ttgering}, \& {Theuns}}]{2008ApJ...681..856R}
{Rauch}, M., {et~al.} 2008, \apj, 681, 856

\bibitem[{{Schaerer}(2003)}]{2003A&A...397..527S}
{Schaerer}, D. 2003, \aap, 397, 527

\bibitem[{{S{\'e}rsic}(1963)}]{1963BAAA....6...41S}
{S{\'e}rsic}, J.~L. 1963, Boletin de la Asociacion Argentina de Astronomia La
  Plata Argentina, 6, 41

\bibitem[{{S{\'e}rsic}(1968)}]{1968adga.book.....S}
---. 1968, {Atlas de Galaxias Australes}

\bibitem[{{Shen} {et~al.}(2003){Shen}, {Mo}, {White}, {Blanton}, {Kauffmann},
  {Voges}, {Brinkmann}, \& {Csabai}}]{2003MNRAS.343..978S}
{Shen}, S., {Mo}, H.~J., {White}, S.~D.~M., {Blanton}, M.~R., {Kauffmann}, G.,
  {Voges}, W., {Brinkmann}, J., \& {Csabai}, I. 2003, \mnras, 343, 978

\bibitem[{{Shibuya} {et~al.}(2012){Shibuya}, {Kashikawa}, {Ota}, {Iye},
  {Ouchi}, {Furusawa}, {Shimasaku}, \& {Hattori}}]{2012ApJ...752..114S}
{Shibuya}, T., {Kashikawa}, N., {Ota}, K., {Iye}, M., {Ouchi}, M., {Furusawa},
  H., {Shimasaku}, K., \& {Hattori}, T. 2012, \apj, 752, 114

\bibitem[{{Shibuya} {et~al.}(2015){Shibuya}, {Ouchi}, \&
  {Harikane}}]{2015ApJS..219...15S}
{Shibuya}, T., {Ouchi}, M., \& {Harikane}, Y. 2015, \apjs, 219, 15

\bibitem[{{Shibuya} {et~al.}(2016){Shibuya}, {Ouchi}, {Kubo}, \&
  {Harikane}}]{2016ApJ...821...72S}
{Shibuya}, T., {Ouchi}, M., {Kubo}, M., \& {Harikane}, Y. 2016, \apj, 821, 72

\bibitem[{{Shibuya} {et~al.}(2014{\natexlab{a}}){Shibuya}, {Ouchi}, {Nakajima},
  {Yuma}, {Hashimoto}, {Shimasaku}, {Mori}, \& {Umemura}}]{2014ApJ...785...64S}
{Shibuya}, T., {Ouchi}, M., {Nakajima}, K., {Yuma}, S., {Hashimoto}, T.,
  {Shimasaku}, K., {Mori}, M., \& {Umemura}, M. 2014{\natexlab{a}}, \apj, 785,
  64

\bibitem[{{Shibuya} {et~al.}(2014{\natexlab{b}}){Shibuya}, {Ouchi}, {Nakajima},
  {Hashimoto}, {Ono}, {Rauch}, {Gauthier}, {Shimasaku}, {Goto}, {Mori}, \&
  {Umemura.}}]{2014ApJ...788...74S}
{Shibuya}, T., {et~al.} 2014{\natexlab{b}}, \apj, 788, 74

\bibitem[{{Shimakawa} {et~al.}(2017){Shimakawa}, {Kodama}, {Shibuya},
  {Kashikawa}, {Tanaka}, {Matsuda}, {Tadaki}, {Koyama}, {Hayashi}, {Suzuki}, \&
  {Yamamoto}}]{2017MNRAS.468.1123S}
{Shimakawa}, R., {et~al.} 2017, \mnras, 468, 1123

\bibitem[{{Shioya} {et~al.}(2009){Shioya}, {Taniguchi}, {Sasaki}, {Nagao},
  {Murayama}, {Saito}, {Ideue}, {Nakajima}, {Matsuoka}, {Trump}, {Scoville},
  {Sanders}, {Mobasher}, {Aussel}, {Capak}, {Kartaltepe}, {Koekemoer},
  {Carilli}, {Ellis}, {Garilli}, {Giavalisco}, {Kitzbichler}, {Impey},
  {LeFevre}, {Schinnerer}, \& {Smolcic}}]{2009ApJ...700..899S}
{Shioya}, Y., {et~al.} 2009, \apj, 700, 899

\bibitem[{{Skelton} {et~al.}(2014){Skelton}, {Whitaker}, {Momcheva}, {Brammer},
  {van Dokkum}, {Labb{\'e}}, {Franx}, {van der Wel}, {Bezanson}, {Da Cunha},
  {Fumagalli}, {F{\"o}rster Schreiber}, {Kriek}, {Leja}, {Lundgren}, {Magee},
  {Marchesini}, {Maseda}, {Nelson}, {Oesch}, {Pacifici}, {Patel}, {Price},
  {Rix}, {Tal}, {Wake}, \& {Wuyts}}]{2014ApJS..214...24S}
{Skelton}, R.~E., {et~al.} 2014, \apjs, 214, 24

\bibitem[{{Smit} {et~al.}(2015){Smit}, {Bouwens}, {Franx}, {Oesch}, {Ashby},
  {Willner}, {Labb{\'e}}, {Holwerda}, {Fazio}, \&
  {Huang}}]{2015ApJ...801..122S}
{Smit}, R., {et~al.} 2015, \apj, 801, 122

\bibitem[{{Sobral} {et~al.}(2017{\natexlab{a}}){Sobral}, {Santos}, {Matthee},
  {Paulino-Afonso}, {Ribeiro}, {Calhau}, \& {Khostovan}}]{2017arXiv171204451S}
{Sobral}, D., {Santos}, S., {Matthee}, J., {Paulino-Afonso}, A., {Ribeiro}, B.,
  {Calhau}, J., \& {Khostovan}, A.~A. 2017{\natexlab{a}}, ArXiv e-prints

\bibitem[{{Sobral} {et~al.}(2017{\natexlab{b}}){Sobral}, {Matthee}, {Best},
  {Stroe}, {R{\"o}ttgering}, {Oteo}, {Smail}, {Morabito}, \&
  {Paulino-Afonso}}]{2017MNRAS.466.1242S}
{Sobral}, D., {et~al.} 2017{\natexlab{b}}, \mnras, 466, 1242

\bibitem[{{Somerville} {et~al.}(2018){Somerville}, {Behroozi}, {Pandya},
  {Dekel}, {Faber}, {Fontana}, {Koekemoer}, {Koo}, {P{\'e}rez-Gonz{\'a}lez},
  {Primack}, {Santini}, {Taylor}, \& {van der Wel}}]{2018MNRAS.473.2714S}
{Somerville}, R.~S., {et~al.} 2018, \mnras, 473, 2714

\bibitem[{{Steidel} {et~al.}(1999){Steidel}, {Adelberger}, {Giavalisco},
  {Dickinson}, \& {Pettini}}]{1999ApJ...519....1S}
{Steidel}, C.~C., {Adelberger}, K.~L., {Giavalisco}, M., {Dickinson}, M., \&
  {Pettini}, M. 1999, \apj, 519, 1

\bibitem[{{Steidel} {et~al.}(2011){Steidel}, {Bogosavljevi{\'c}}, {Shapley},
  {Kollmeier}, {Reddy}, {Erb}, \& {Pettini}}]{2011ApJ...736..160S}
{Steidel}, C.~C., {Bogosavljevi{\'c}}, M., {Shapley}, A.~E., {Kollmeier},
  J.~A., {Reddy}, N.~A., {Erb}, D.~K., \& {Pettini}, M. 2011, \apj, 736, 160

\bibitem[{{Taniguchi} {et~al.}(2009){Taniguchi}, {Murayama}, {Scoville},
  {Sasaki}, {Nagao}, {Shioya}, {Saito}, {Ideue}, {Nakajima}, {Matsuoka},
  {Sanders}, {Mobasher}, {Aussel}, {Capak}, {Salvato}, {Koekemoer}, {Carilli},
  {Cimatti}, {Ellis}, {Garilli}, {Giavalisco}, {Ilbert}, {Impey},
  {Kitzbichler}, {Le F{\`e}vre}, {McCracken}, {Scarlata}, {Schinnerer},
  {Smolcic}, {Tribiano}, \& {Trump}}]{2009ApJ...701..915T}
{Taniguchi}, Y., {et~al.} 2009, \apj, 701, 915

\bibitem[{{van der Wel} {et~al.}(2014){van der Wel}, {Franx}, {van Dokkum},
  {Skelton}, {Momcheva}, {Whitaker}, {Brammer}, {Bell}, {Rix}, {Wuyts},
  {Ferguson}, {Holden}, {Barro}, {Koekemoer}, {Chang}, {McGrath},
  {H{\"a}ussler}, {Dekel}, {Behroozi}, {Fumagalli}, {Leja}, {Lundgren},
  {Maseda}, {Nelson}, {Wake}, {Patel}, {Labb{\'e}}, {Faber}, {Grogin}, \&
  {Kocevski}}]{2014ApJ...788...28V}
{van der Wel}, A., {et~al.} 2014, \apj, 788, 28

\bibitem[{{Venemans} {et~al.}(2005){Venemans}, {R{\"o}ttgering}, {Miley},
  {Kurk}, {De Breuck}, {Overzier}, {van Breugel}, {Carilli}, {Ford}, {Heckman},
  {Pentericci}, \& {McCarthy}}]{2005AA...431..793V}
{Venemans}, B.~P., {et~al.} 2005, \aap, 431, 793

\bibitem[{{Vitvitska} {et~al.}(2002){Vitvitska}, {Klypin}, {Kravtsov},
  {Wechsler}, {Primack}, \& {Bullock}}]{2002ApJ...581..799V}
{Vitvitska}, M., {Klypin}, A.~A., {Kravtsov}, A.~V., {Wechsler}, R.~H.,
  {Primack}, J.~R., \& {Bullock}, J.~S. 2002, \apj, 581, 799

\bibitem[{{Warren} {et~al.}(1992){Warren}, {Quinn}, {Salmon}, \&
  {Zurek}}]{1992ApJ...399..405W}
{Warren}, M.~S., {Quinn}, P.~J., {Salmon}, J.~K., \& {Zurek}, W.~H. 1992, \apj,
  399, 405

\bibitem[{{Wisotzki} {et~al.}(2016){Wisotzki}, {Bacon}, {Blaizot},
  {Brinchmann}, {Herenz}, {Schaye}, {Bouch{\'e}}, {Cantalupo}, {Contini},
  {Carollo}, {Caruana}, {Courbot}, {Emsellem}, {Kamann}, {Kerutt}, {Leclercq},
  {Lilly}, {Patr{\'{\i}}cio}, {Sandin}, {Steinmetz}, {Straka}, {Urrutia},
  {Verhamme}, {Weilbacher}, \& {Wendt}}]{2016AA...587A..98W}
{Wisotzki}, L., {et~al.} 2016, \aap, 587, A98

\bibitem[{{Wold} {et~al.}(2014){Wold}, {Barger}, \&
  {Cowie}}]{2014ApJ...783..119W}
{Wold}, I.~G.~B., {Barger}, A.~J., \& {Cowie}, L.~L. 2014, \apj, 783, 119

\bibitem[{{Wold} {et~al.}(2017){Wold}, {Finkelstein}, {Barger}, {Cowie}, \&
  {Rosenwasser}}]{2017ApJ...848..108W}
{Wold}, I.~G.~B., {Finkelstein}, S.~L., {Barger}, A.~J., {Cowie}, L.~L., \&
  {Rosenwasser}, B. 2017, \apj, 848, 108

\bibitem[{{Xu} {et~al.}(2007){Xu}, {Pirzkal}, {Malhotra}, {Rhoads}, {Mobasher},
  {Daddi}, {Gronwall}, {Hathi}, {Panagia}, {Ferguson}, {Koekemoer},
  {K{\"u}mmel}, {Moustakas}, {Pasquali}, {di Serego Alighieri}, {Vernet},
  {Walsh}, {Windhorst}, \& {Yan}}]{2007AJ....134..169X}
{Xu}, C., {et~al.} 2007, \aj, 134, 169

\bibitem[{{Xue} {et~al.}(2017){Xue}, {Lee}, {Dey}, {Reddy}, {Hong}, {Prescott},
  {Inami}, {Jannuzi}, \& {Gonzalez}}]{2017ApJ...837..172X}
{Xue}, R., {et~al.} 2017, \apj, 837, 172

\bibitem[{{Yang} {et~al.}(2017{\natexlab{a}}){Yang}, {Malhotra}, {Rhoads},
  {Leitherer}, {Wofford}, {Jiang}, \& {Wang}}]{2017ApJ...838....4Y}
{Yang}, H., {Malhotra}, S., {Rhoads}, J.~E., {Leitherer}, C., {Wofford}, A.,
  {Jiang}, T., \& {Wang}, J. 2017{\natexlab{a}}, \apj, 838, 4

\bibitem[{{Yang} {et~al.}(2017{\natexlab{b}}){Yang}, {Malhotra}, {Gronke},
  {Rhoads}, {Leitherer}, {Wofford}, {Jiang}, {Dijkstra}, {Tilvi}, \&
  {Wang}}]{2017ApJ...844..171Y}
{Yang}, H., {et~al.} 2017{\natexlab{b}}, \apj, 844, 171

\bibitem[{{Zheng} \& {Wallace}(2013)}]{2013arXiv1308.1405Z}
{Zheng}, Z., \& {Wallace}, J. 2013, ArXiv e-prints

\bibitem[{{Zheng} {et~al.}(2017){Zheng}, {Wang}, {Rhoads}, {Infante},
  {Malhotra}, {Hu}, {Walker}, {Jiang}, {Jiang}, {Hibon}, {Gonzalez}, {Kong},
  {Zheng}, {Galaz}, \& {Barrientos}}]{2017ApJ...842L..22Z}
{Zheng}, Z.-Y., {et~al.} 2017, \apjl, 842, L22

\end{thebibliography}

\end{document}